\documentclass[twocolumn,aps,prb,showpacs]{revtex4-1}
\usepackage[utf8]{inputenc}
\usepackage[normalem]{ulem}
\usepackage[T1]{fontenc}
\usepackage[unicode=true, pdfusetitle, bookmarks=true, bookmarksnumbered=false, bookmarksopen=false, breaklinks=false, pdfborder={0 0 1}, backref=false, colorlinks=false]{hyperref}
\hypersetup{colorlinks,linkcolor=blue,citecolor=blue,urlcolor=blue}
\usepackage{float}
\usepackage{verbatim}
\usepackage[abs]{overpic}
\usepackage{color}
\usepackage{comment}
\usepackage{amsmath,amssymb}
\usepackage{graphicx}

\begin{document}

\title{Entanglement and boundary entropy in quantum spin chains with arbitrary direction of the boundary magnetic fields}

\author{J. C. Xavier}
\affiliation{Universidade Federal de Uberl\^andia, Instituto de F\'isica, C. P 593, 38400-902 Uberl\^andia, MG, Brazil}
\author{ M.~A.~Rajabpour}
\affiliation{  Instituto de F\'isica, Universidade Federal Fluminense, Av. Gal. Milton Tavares de Souza s/n, Gragoat\'a, 24210-346, Niter\'oi, RJ, Brazil}
\date{\today{}}

\begin{abstract}
We calculate the entanglement and the universal boundary entropy (BE) in the critical quantum spin chains, such as the transverse field Ising chain and the XXZ chain, with arbitrary direction of the boundary magnetic field (ADBMF). We determine the boundary universality class that an ADBMF induces. In particular, we show that the induced boundary conformal field theory (BCFT) depends on the point on the Bloch sphere where the boundary magnetic field directs. We show that the classification of the directions boils down to the simple fact that the boundary field breaks the bulk symmetry or does not. We present a procedure to estimate the universal BE, based on the finite-size corrections of the entanglement entropy, that apply to  the ADBMF. To calculate the universal BE in the XXZ chain, we use  the density matrix renormalization group (DMRG).  The transverse field XY chain with ADBMF after Jordan-Wigner (JW) transformation is not a quadratic free fermion Hamiltonian. We map this model to a quadratic free fermion chain
by introducing two extra ancillary spins coupled to the main chain at the boundaries, which makes the problem {\it{integrable}}. The eigenstates of the transverse field XY chain can be obtained  by  proper projection in the enlarged chain. Using this mapping, we are able to calculate the entanglement entropy of the transverse field XY chain using the usual correlation matrix technique up to relatively large sizes. 
\end{abstract}

\maketitle

\section{Introduction}
\label{sec:intro}

Quantum entanglement in many body systems has been studied in a great detail
in the last couple of decades. There are comprehensive reviews on the
applications of the entanglement entropy in condensed matter physics
\cite{laflorencie2016quantum}, quantum field theories \cite{Casini-2009},
integrable models \cite{castro2009bi} and conformal field theory (CFT)
\cite{calabrese2009entanglement}. There are several motivations to study
quantum entanglement. For instance, the entanglement is an essential
ingredient in quantum computation and in many other applications\cite{laflorencie2016quantum,Horodeckirev}.
For these reasons, the quantification of the entanglement have been studied
extensively. The entanglement entropy, which is one of the most used quantifier of entanglement
of pure bipartite systems, was measured recently in one-dimensional quantum systems \cite{OpticalRenyi-1,OpticalRenyi-2}.

The bipartite entanglement entropy is defined as $S=-\text{tr}{\rho_A\ln\rho_A}$,
where $\rho_A$ is the reduced density matrix of the subsystem $A$. It has been well understood for the ground state of critical and non-critical quantum chains. One of the great outcomes of all these studies was the central role of the entanglement entropy to distinguish different phases and classify the critical point of the continuous phase transitions to different universality classes consistent with the traditional classification based on local observables. Most of the above studies were based on the bulk properties, however, there are also many studies regarding the entanglement entropy in systems with  boundaries. In the presence of boundaries analytical and numerical calculations of the entanglement entropy is normally a bit more challenging because of the lack of the translational invariance. Nevertheless, the entanglement entropy of a few quantum chains in the presence of the boundaries has been studied with analytical and numerical techniques, see for instance Refs. \onlinecite{Laflorencie-boundary2006,Barthel2006b,Legeza2007,Szirmai2008,Castro_Alvaredo_2009,Affleck2009,Taddia2013,Fagotti2011} and Ref. \onlinecite{laflorencie2016quantum} for a review.

In the presence of a boundary there is an interesting degree of freedom in which the bulk of the system can be at critical point, but the boundary can be non-critical and flow between different fixed points under boundary renormalization group, see Ref. \onlinecite{Diehl1997} and references therein. In one spatial dimension such kind of flow in the language of CFT in connection with the impurity problems such as the Kondo problem, has already been studied \cite{Affleck1991}. The interesting observation of Ref. \onlinecite{Affleck1991} is that for a system with its bulk at the critical point one can define a BE which decreases under boundary renormalization group and at the boundary fixed point is equal to a number which is related to the universality class of the corresponding boundary condition. This BE in the context of the entanglement entropy has been studied in CFT \cite{calabrese2004entanglement,Najafi-2016,Alba2017}, quantum spin chains \cite{Laflorencie-boundary2006,Barthel2006b,Legeza2007,Szirmai2008,Taddia2013,Fagotti2011,Tu2017,Tang2017,Najafi-2016} and integrable models \cite{Castro_Alvaredo_2009}. In particular, using DMRG technique the authors of Ref. \onlinecite{Barthel2006}  estimate the universal BE for the transverse field Ising  chain with particular boundary conditions, mainly boundary magnetic field in the $x$ direction. In this work, we would like to generalize this idea in a few directions.

From the physics point of view, it is interesting to classify  the boundary conditions in the quantum spin chains when there is a magnetic field at the boundary in an arbitrary direction. We would like to do this classification by evaluating the contribution of BE. The precise determination of the BE, even numerically, can be challenging from the technical point of view. In this vein, we present a simple procedure to estimate the BE which is based on the finite-size scaling of the entanglement entropy.
In order to illustrate that the procedure works quite well, we consider the two most interesting models: the transverse field Ising chain and the XXZ chain. For the XY chain (the transverse field Ising chain is a particular case of this model) in the presence of the ADBMF, we were able to map the problem of the diagonalization of the model
to a free fermion model which can be diagonalized in a linear time. To the
best of our knowledge this problem has not been tackled before in the
literature (see Ref. \onlinecite{Bilstein1999} for the solution of the XX
chain with ADBMF) and it seems interesting for its own sake. After solving the
XY chain with ADBMF we use a modified version of the Peschel method
\cite{peschel2003calculation} to calculate the entanglement entropy of finite
systems. In the case of the XXZ chain with ADBMF we tackle the problem with
the DMRG \cite{white,reviewdmrgS}. Having the the finite-size corrections of
the entanglement entropy of those models, we show that it is possible to
estimate the BE for ADBMF and classify the corresponding boundary CFT, using
relatively large system sizes.

The paper is organized as follows: In the section \ref{sec:BE-CFT}, we first define the universal BE and introduce the relevant equations and notations. In  section \ref{sec:XY}, we solved the Hamiltonian of the XY chain with ADBMF. Then we find the correlation functions and generalize the Peschel method to calculate the entanglement entropy. We close this section  presenting our numerical results regarding the universal BE in the transverse field Ising chain with ADBMF.
In section \ref{sec:XXZ}, we study the boundary entanglement entropy in the XXZ chain with ADBMF using DMRG technique. Finally, in section \ref{sec:conc}, we summarize our findings.

\section{Boundary entropy in CFT}
\label{sec:BE-CFT}
Consider a one-dimensional quantum field theory defined on a system of length $L$ with boundary conditions $a$ and $b$ at $x=0$ and $L$, respectively. The partition function of this system at inverse temperature $\beta$ is given by the partition function of a two dimensional system on a cylinder with particular BCs. According to Affleck-Ludwig argument \cite{Affleck1991} when we have a CFT in the limit of large $L$ and $\beta$ the free energy of the system behaves as $F=-Lf+f'$, where $f'$ has in addition to the non-universal contribution also a universal BE, i.e. $f_a+f_b$, with $f_{a,b}=-\frac{1}{\beta}\ln g_{a,b}$, where $g_{a}=\langle B_a|0\rangle$ and $g_{b}=\langle 0|B_b\rangle$. The states $|B_{a,b}\rangle$  are  the so-called conformal boundary states \cite{CARDY1989,CARDY1991}. This means that one-dimensional CFT defined on a segment has a non trivial zero temperature BE, or ground state degeneracy. In other words, we have a ground state degeneracy which should be understood as particular behavior of the low-energy density of states in CFT. It was suggested in \cite{Affleck1991} and later proved in \cite{Friedan2004} that the quantity $g$ decreases under the boundary renormalization group for systems that the bulk is already at the critical phase. 

The equivalent description using the entanglement entropy was first suggested in \cite{calabrese2004entanglement}. The idea goes as follows: 
Consider first the entanglement entropy of the ground state of a CFT with periodic boundary condition (PBC) and length $L$, then the entanglement entropy of a subsystem with length $\ell$ behaves as \cite{HOLZHEY1994443,calabrese2004entanglement,Vidal2001,Korepin2004}
\begin{equation}\label{EEpbc}
S^{PBC}(L,\ell)=\frac{c}{3}\ln\left[\frac{L}{\pi}\sin\left(\frac{\pi \ell}{L}\right)\right]+c_{1}^{PBC},
\end{equation}
 where $c$ is the central charge. 
 However, for systems with boundaries the entanglement entropy of a segment starting from one boundary and with length $\ell$ behaves as    \cite{calabrese2004entanglement,Barthel2006,Affleck2009,Taddia2013}
\begin{equation}\label{EEb}
S^{b}(L,\ell)=\frac{c}{6}\ln\left[\frac{2L}{\pi}\sin\left(\frac{\pi \ell}{L}\right)\right]+c_{1}^{PBC}/2+\ln(g)+G_{b}(\frac{\ell}{L})\; ,
\end{equation}
where $s_{b}=\ln(g)$ is the BE and  $G_{b}(x)=\lim_{n\rightarrow1}\frac{1}{1-n}\ln{F_{\Upsilon}^{(n)}}$ where

\begin{equation}
 F_\Upsilon^{(n)}(x)=\frac{e^{i2\pi(n-1)h_\Upsilon}}{n^{2nh_\Upsilon}}\frac{\left<\prod_{k=0}^{n-1}\Upsilon(z_{n,k}^-)\Upsilon^\dagger(z_{n,k}^+)\right>}{\left<\Upsilon(z_{1,0}^-)\Upsilon^\dagger(z_{1,0}^+)\right>^n}, 
 \label{fcom}
 \end{equation}
$z_{k, n}^\pm = e^{ \frac{ i \pi}{n} ( \pm  x + 2 k)},  \ k=0,1,  \dots, n-1,$ and  $\Upsilon$ is a chiral primary field with conformal dimension $h_\Upsilon$ \cite{Taddia2013}.

 We intent to get estimates of the BE, or equivalently the ground state degeneracy $g$, by using the finite-size scaling of the entanglement entropies introduced in the above. When $G_b=0$, this task is quite simple. First, we get the non-universal constant $c_1^{PBC}$ by using Eq. (\ref{EEpbc}) and plug this result in Eq. (\ref{EEb}) to obtain $s_b$.  The authors of the Ref. \onlinecite{Barthel2006} used this procedure to extract $s_b$  of the transverse field Ising chain, with open boundary condition (OBC) and boundary magnetic fields in the $x$ direction. However, in general, $G_b$ is non-zero and if we do not know this function, we can not use this procedure to determine $s_b$. Of course {\it if } we knew the non-universal function $G_b$, in principle, we could use the same procedure to extract $s_b$.  It is important to emphasize that obtaining $G_b$, or equivalently  $F_\Upsilon^{(n)}(x)$ which depends on the model, is not a simple task. For integer values of $n$,   $F_\Upsilon^{(n)}(x)$ is known for the free boson
theory as well as the transverse field Ising chain, see Refs. \onlinecite{Alcaraz2011} and \onlinecite{Taddia2013}. Note that to calculate $G_b$ it is necessary to analytically continue the function $F_\Upsilon^{(n)}(x)$ to non-integers values of $n$ which  is a non-trivial task, see for instance Ref. \onlinecite{Esslerfunction}.

Motivated by the above discussion, we present a procedure to obtain $s_{b}$, numerically, {\it without knowing} the function $G_b$, as we explain in the following. First, note that BE is related with the entranglement entropies of systems under PBC and with boundaries by   
\begin{equation}\label{sb}
s_{b}=\ln(g)=S^{b}-\frac{S^{PBC}}{2}-\frac{c}{6}\ln(2)-G_b(\frac{\ell}{L})\;.
\end{equation}
Due to the finite-size scaling of the entanglement entropies defined in Eqs. (\ref{EEpbc}) and (\ref{EEb})
it is convenient to define $f(x)=S^{b}-\frac{S^{PBC}}{2}-\frac{c}{6}\ln(2)=\ln(g)+G_b(\frac{\ell}{L})$. Now, inspired by the behavior of the non-universal function $G_b(x)$ for small values of $x$ (see Ref. \onlinecite{Alcaraz2011}),  we assume  $f(x)$ behaves as 
\begin{equation}
f(x)=\ln(g)+\frac{2\pi^{2}}{3}a_{1}x^{2}+a_{2}x^{4}+a_{3}/x^{a_{4}},\label{fx}
\end{equation}
where we have added the term $ a_3/x^{a_{4}}$ due to the unusual
corrections \citep{Campostrini2010,Cardy2010,Xavier2012,Ercolessi2012,Dalmonte2011}. So, to get numerical estimates of the BE we  fit the numerical data of the entranglement entropies to Eq. (\ref{fx}).
We will illustrate the above procedure for the transverse field Ising chain and the XXZ model with ADBMF in the next sections. 

We close this section mentioning that in order to investigate which are the conformally invariant boundary conditions, we apply boundary magnetic fields $h_b$ in the lattice models. Usually, for $0<h_b<\infty$ the conformal invariance is lost and we can associate a crossover length $\xi \sim h_b^{1-d}$, where $d<1$ is the scaling dimension of the relevant boundary perturbation. The above equation is valid only for $\ell>\xi$ \cite{affleckboundaryxxz}.

\section{The Transverse field XY chain with arbitrary direction of the boundary magnetic field }
\label{sec:XY}
In this section, we study the entanglement entropy and universal BE in the transverse field XY chain. To calculate the entanglement entropy we first solve the transverse field XY chain with ADBMF. Since this problem is interesting for its own sake and to the best of our knowledge has not been investigated in its full generality, we provide some details regarding its solution. After finding the ground state we  show how one can find the entanglement entropy using the correlation matrix method. Here too, there are a few new subtleties that should be addressed and for that reason we provide some detailed calculations. The advantage of the correlation matrix method is that its complexity grows linearly with the system size while for other methods usually grows exponentially.
Finally, we use our refined correlation method to calculate the entanglement entropy and the BE.

\subsection{Diagonalization of the finite chain using the ghost site technique}
\label{subsec:diagonalization}
In this section, we show how it is possible to obtain the energies and the correlation matrices of the transverse field XY chain in the presence of ADBMF. The route we are going to follow is the same as the one used in Ref. \onlinecite{Bilstein1999} for the XX chain with boundary magnetic fields.  For earlier use of the same technique see Refs. \onlinecite{Colpa_1979,BARIEV1991,Campostrini2015}. We consider the following XY Hamiltonian with ADBMF  
\begin{eqnarray}
\ \textbf{H}^{XY}=J\sum_{j=1}^{L-1}\Big{[}(1+\gamma)S_{j}^{x}S_{j+1}^{x}+(1-\gamma)S_{j}^{y}S_{j+1}^{y}\Big{]}\nonumber\\
-h\sum_{j=1}^{L}S_{j}^{z}+\vec{b}_{1}\cdot\vec{S}_{1}+\vec{b}_{L}\cdot\vec{S}_{L}\,,\hspace{1.8cm}\label{HXY}
\end{eqnarray}
 where $S^{\alpha}=\frac{1}{2}\sigma^{\alpha}$, $\sigma^{\alpha}$($\alpha=x,y,z)$
are the Pauli matrices and 
\begin{eqnarray}
\ \vec{b}_{1} & = & b_{1}(\sin\theta_{1}\cos\phi_{1},\sin\theta_{1}\sin\phi_{1},\cos\theta_{1}),\label{b1}\\
\vec{b}_{L} & = & b_{L}(\sin\theta_{L}\cos\phi_{L},\sin\theta_{L}\sin\phi_{L},\cos\theta_{L})\,.\label{b2}
\end{eqnarray}
The transverse field XY chain has a few interesting critical lines. On $h=\pm J$ and $\gamma\neq0$ we have the universality of the critical Ising chain with the central charge $c=\frac{1}{2}$. On $\gamma=0$ and $-1<h<1$ we have the universality class of the compactified bosons with the central charge $c=1$. The other parts of the phase diagram are not critical and although all the calculations of this section are valid for the full phase diagram, in the later sections we will concentrate mostly on the critical parts of the phase diagram.

 It is convenient to rewrite the above Hamiltonian in terms of $\sigma^{\pm}=\frac{\sigma^{x}\pm i\sigma^{y}}{2}$
and $\sigma^{z}$ as follows:
\begin{eqnarray}
\ \textbf{H}^{XY} & = & \frac{J}{2}\sum_{j=1}^{L-1}\Big{(}\sigma_{j}^{+}\sigma_{j+1}^{-}+\gamma\sigma_{j}^{+}\sigma_{j+1}^{+}+h.c.\Big{)}-\frac{h}{2}\sum_{j=1}^{L}\sigma_{j}^{z}+\nonumber \\
 & + & \frac{1}{2}b_{1}\Big{(}\sin\theta_{1}e^{-i\phi_{1}}\sigma_{1}^{+}+\sin\theta_{1}e^{i\phi_{1}}\sigma_{1}^{-}+\cos\theta_{1}\sigma_{1}^{z}\Big{)}\nonumber \\
 & + & \frac{1}{2}b_{L}\Big{(}\sin\theta_{L}e^{-i\phi_{L}}\sigma_{L}^{+}+\sin\theta_{L}e^{i\phi_{L}}\sigma_{L}^{-}+\cos\theta_{L}\sigma_{L}^{z}\Big{)}\,.\label{HXYsig}\nonumber
\end{eqnarray}

Note that if we try to diagonalize the above Hamiltonian by using
the Jordan-Wigner transformation we realize that the fermionic Hamiltonian
is not in a bilinear form in terms of creation and annihilation operators.
In order to circumvent this issue, we follow the procedure used in Ref. \onlinecite{Bilstein1999} and consider another spin Hamiltonian, $\textbf{H}^{long}$, which
has two extra sites $(0$ and $L+1)$, i.e. ghost sites, interacting with the the spins
at sites $1$ and $L$, as explained in the following. We first define the following enlarged Hamiltonian\begin{widetext}
\begin{eqnarray}
\  & \textbf{H}^{long}= & \frac{J}{2}\sum_{j=1}^{L-1}\Big{[}(\sigma_{j}^{+}\sigma_{j+1}^{-}+\sigma_{j}^{-}\sigma_{j+1}^{+})+\gamma(\sigma_{j}^{+}\sigma_{j+1}^{+}+\sigma_{j}^{-}\sigma_{j+1}^{-})\Big{]}-\frac{h}{2}\sum_{j=1}^{L}\sigma_{j}^{z}+\nonumber \\
 & + & \frac{1}{2}b_{1}\Big{(}\sin\theta_{1}e^{-i\phi_{1}}\sigma_{0}^{x}\sigma_{1}^{+}+\sin\theta_{1}e^{i\phi_{1}}\sigma_{0}^{x}\sigma_{1}^{-}+\cos\theta_{1}\sigma_{1}^{z}\Big{)}\nonumber \\
 & + & \frac{1}{2}b_{L}\Big{(}\sin\theta_{L}e^{-i\phi_{L}}\sigma_{L}^{+}\sigma_{L+1}^{x}+\sin\theta_{L}e^{i\phi_{L}}\sigma_{L}^{-}\sigma_{L+1}^{x}+\cos\theta_{L}\sigma_{L}^{z}\Big{)}\,.\label{Hlong}
\end{eqnarray}
\end{widetext}
The above Hamiltonian commutes with $\sigma_{0}^{x}$ and $\sigma_{L+1}^{x}$.
Due to this fact it is possible to block diagonalize $\textbf{H}^{long}$
in four distinct sectors, labeled by the eigenvalues of $\sigma_{0}^{x}$
and $\sigma_{L+1}^{x}$. We are going to denote these sectors by
$(s_{0},s_{L+1})$, where $s_{j}=\pm1$ ($j=0$ or $j=L+1$) are the
eigenvalues of $\sigma_{j}^{x}$ whose eigenstates are $\mid s_{j}\rangle=\frac{1}{\sqrt{2}}(\mid\uparrow\rangle+s_{j}\mid\downarrow\rangle)$\,.  
Note that 
\begin{equation}
\mid\Psi_{k}^{long}(+,+)\rangle=\mid+\rangle\otimes\mid\Psi_{k}^{XY}\rangle\otimes\mid+\rangle\,,\label{stateHlong}
\end{equation}
are eigenstates of $\textbf{H}^{long}$ and $\mid\Psi_{k}^{XY}\rangle$ are the eigenstates of the XY chain with energy $E_{k}^{XY}$. 
This means that by  knowing $\mid\Psi_{k}^{long}(+,+)\rangle$ it is possible to obtain
$\mid\Psi_{k}^{XY}\rangle$ by projecting, appropriately, the state $\mid\Psi_{k}^{long}(+,+)\rangle$. This procedure will be discussed in detail below. It is worth mentioning that the spectrum of $\textbf{H}^{long}$ will be
at least \emph{twice degenerate} due to the fact that $\textbf{H}^{long}$
is invariant under the transformation $\sigma_{l}^{\alpha}\rightarrow-\sigma_{l}^{\alpha}$
with $\alpha=x,y$ and $\sigma_{l}^{z}\rightarrow\sigma_{l}^{z}$.

Using the Jordan-Wigner transformation 
\begin{eqnarray}
\ c_{l}^{\dagger}=\prod_{j=0}^{l-1}\sigma_{j}^{z}\sigma_{l}^{+},\hspace{1cm}c_{l}=\prod_{j=0}^{l-1}\sigma_{j}^{z}\sigma_{l}^{-},\label{Jordan-Wigner}
\end{eqnarray}
we map the Hamiltonian $\textbf{H}^{long}$ to the following free
fermion Hamiltonian 
\begin{widetext}
\begin{eqnarray}
\ \textbf{H}_{ff}^{long}=-\frac{J}{2}\sum_{j=1}^{L-1}\Big{(}c_{j}^{\dagger}c_{j+1}+\gamma c_{j}^{\dagger}c_{j+1}^{\dagger}+c_{j+1}^{\dagger}c_{j}+\gamma c_{j+1}c_{j}\Big{)}-\frac{h}{2}\sum_{j=1}^{L}(2c_{j}^{\dagger}c_{j}-1)\hspace{8cm}\nonumber \\
-\frac{b_{1}}{2}\Big{[}\sin\theta_{1}e^{i\phi_{1}}c_{0}^{\dagger}c_{1}+\sin\theta_{1}e^{-i\phi_{1}}c_{0}^{\dagger}c_{1}^{\dagger}+\sin\theta_{1}e^{-i\phi_{1}}c_{1}^{\dagger}c_{0}+\sin\theta_{1}e^{i\phi_{1}}c_{1}c_{0}-\cos\theta_{1}(2c_{1}^{\dagger}c_{1}-1)\Big{]}\hspace{4.5cm}\nonumber \\
-\frac{b_{L}}{2}\Big{[}\sin\theta_{L}e^{i\phi_{L}}c_{L+1}^{\dagger}c_{L}+\sin\theta_{L}e^{-i\phi_{L}}c_{L}^{\dagger}c_{L+1}^{\dagger}+\sin\theta_{L}e^{-i\phi_{L}}c_{L}^{\dagger}c_{L+1}+\sin\theta_{L}e^{i\phi_{L}}c_{L+1}c_{L}-\cos\theta_{L}(2c_{L}^{\dagger}c_{L}-1)\Big{]}, \label{Hlongfree}\hspace{2cm}
\end{eqnarray}
\end{widetext}
which is bilinear in terms of creation and annihilation operators
and can be diagonalized using the standard method that will be explained in the following. It is convenient to write the above Hamiltonian as
\begin{eqnarray}
\ \textbf{H}_{ff}^{long}=\sum_{i,j=0}^{L+1}[c_{i}^{\dagger}A_{ij}c_{j}+\frac{1}{2}c_{i}^{\dagger}B_{ij}c_{j}^{\dagger}+\frac{1}{2}c_{i}B_{ji}^{*}c_{j}]-\frac{1}{2}{\rm Tr}{\textbf{A}^{*}}\label{Hfreefermion}\nonumber
\end{eqnarray}
where the exact forms of the matrices $\textbf{A}$ and $\textbf{B}$ which are crucial for later calculations and arguments are
\begin{widetext}
\begin{eqnarray}
\ \textbf{A}=\begin{pmatrix}0 & -\frac{1}{2}b_{1}\sin\theta_{1}e^{i\phi_{1}} & 0 & ...\\
-\frac{1}{2}b_{1}\sin\theta_{1}e^{-i\phi_{1}} & -h+b_{1}\cos\theta_{1} & -\frac{J}{2} & 0 & ...\\
0 & -\frac{J}{2} & -h & -\frac{J}{2} & 0 & ...\\
0 & 0 & -\frac{J}{2} & -h & -\frac{J}{2} & 0 & ...\\
. & . & .\\
. & . & .\\
. & . & . & . & . & . & -h+b_{L}\cos\theta_{L} & -\frac{1}{2}b_{L}\sin\theta_{L}e^{-i\phi_{L}}\\
. & . & . & . & . & . & -\frac{1}{2}b_{L}\sin\theta_{L}e^{i\phi_{L}} & 0
\end{pmatrix},\label{matrixA}
\end{eqnarray}
\end{widetext}
and 
\begin{widetext}
\begin{eqnarray}
\ \textbf{B}=\begin{pmatrix}0 & -\frac{1}{2}b_{1}\sin\theta_{1}e^{-i\phi_{1}} & 0 & ...\\
\frac{1}{2}b_{1}\sin\theta_{1}e^{-i\phi_{1}} & 0 & -\frac{\gamma J}{2} & 0 & ...\\
0 & \frac{\gamma J}{2} & 0 & -\frac{\gamma J}{2} & 0 & ...\\
0 & 0 & \frac{\gamma J}{2} & 0 & -\frac{\gamma J}{2} & 0 & ...\\
. & . & .\\
. & . & .\\
. & . & . & . & . & . & \frac{\gamma J}{2} & 0 & -\frac{1}{2}b_{L}\sin\theta_{L}e^{-i\phi_{L}}\\
. & . & . & . & . & . & 0 & \frac{1}{2}b_{L}\sin\theta_{L}e^{-i\phi_{L}} & 0
\end{pmatrix}.\hspace{1cm}\label{matrixB}
\end{eqnarray}
\end{widetext}
Note that the matrix $\textbf{A}$ is Hermitian while the matrix $\textbf{B}$
is antisymmetric. 

In order to diagonalize the above Hamiltonian, it is convenient to
rewrite it in the following matrix form 
\begin{eqnarray}
\ \textbf{H}_{ff}^{long}=\frac{1}{2}(\textbf{c}^{\dagger}\,\,\,\textbf{c})\textbf{M}\begin{pmatrix}\textbf{c}\\
\textbf{c}^{\dagger}
\end{pmatrix},\label{HfreefermionM}
\end{eqnarray}
where 
\begin{eqnarray}
\ \textbf{M}=\begin{pmatrix}\textbf{A} & \textbf{B}\\
-\textbf{B}^{*} & -\textbf{A}^{*}
\end{pmatrix},\label{matrixM}
\end{eqnarray}
is a $(2L+4)\times(2L+4)$ Hermitian matrix and we denote $(\textbf{c}^{\dagger}\,\,\,\textbf{c})=(c_{0}^{\dagger},c_{1}^{\dagger},\dots,c_{L+1}^{\dagger},c_{0},c_{1},\dots,c_{L+1})$. 

Due to the especial form of the Hermitian matrix $\textbf{M}$ one can always find a unitary matrix in the form
\begin{eqnarray}
\ \textbf{U}=\begin{pmatrix}\boldsymbol{g} & \boldsymbol{h}\\
\boldsymbol{h}^{*} & \boldsymbol{g}^{*}
\end{pmatrix},\label{Ud}
\end{eqnarray}
which diagonalizes the matrix $\textbf{M}$, where $\boldsymbol{g}$ and $\boldsymbol{h}$ are $(L+2)\times(L+2)$ matrices. Due to this fact,  one can write
\begin{eqnarray}
\ \textbf{H}_{ff}^{long} & = & \frac{1}{2}(\textbf{c}^{\dagger}\,\,\,\textbf{c})\textbf{U}^{\dagger}\textbf{U}\begin{pmatrix}\textbf{A} & \textbf{B}\\
-\textbf{B}^{*} & -\textbf{A}^{*}
\end{pmatrix}\textbf{U}^{\dagger}\textbf{U}\begin{pmatrix}\textbf{c}\\
\textbf{c}^{\dagger}
\end{pmatrix},\nonumber \\
 & = & \frac{1}{2}(\boldsymbol{\eta}^{\dagger}\,\,\,\boldsymbol{\eta})\boldsymbol{\Lambda}\begin{pmatrix}\boldsymbol{\eta}\\
\boldsymbol{\eta}^{\dagger}
\end{pmatrix},
\end{eqnarray}
where we introduced new fermionic operators 
\begin{eqnarray}
\ \begin{pmatrix}\boldsymbol{\eta}\\
\boldsymbol{\eta}^{\dagger}
\end{pmatrix}=\textbf{\ensuremath{\textbf{U}}}\begin{pmatrix}\textbf{c}\\
\textbf{c}^{\dagger}
\end{pmatrix}.\label{canonical-transformation}
\end{eqnarray}
Furthermore it is easy to prove that the eigenvalues of the matrix $\textbf{M}$ appear in pairs, i. e.
$\pm\lambda_{i}$ and one can write
\begin{eqnarray}
\ \boldsymbol{\Lambda}=\begin{pmatrix}\boldsymbol{\Lambda}_{1} & \textbf{O}\\
\textbf{O} & -\boldsymbol{\Lambda}_{1}
\end{pmatrix}.\label{lambda}
\end{eqnarray}
Finally, one can write the
diagonalized form of the Hamiltonian as follows: 
\begin{eqnarray}
\ \textbf{H}_{ff}^{long}=\sum_{k}\lambda_{k}\eta_{k}^{\dagger}\eta_{k}-\frac{1}{2}{\rm Tr}{\boldsymbol{\Lambda}_{1}},\label{Hdiagfinal}
\end{eqnarray}
where the modes are ordered as $0=\lambda_{0}\leq\lambda_{1}\leq\cdots\leq\lambda_{L+1}.$

It is easy to check that the matrix $\textbf{M}$ has \emph{at
least two} eigenvectors corresponding to the zero eigenvalue, due
to the form of the matrix $\textbf{A}$ and $\textbf{B}.$ They have
the following forms 
\begin{eqnarray}
\hspace*{-0.2cm}|u_{0}^{1}\rangle=\begin{pmatrix}\sqrt{a}e^{i\alpha_{0}^{1}}\\
0\\
\vdots\\
0\\
\sqrt{\frac{1}{2}-a}e^{i\gamma_{0}^{1}}\\
\sqrt{a}e^{i\alpha_{0}^{1}}\\
0\\
\vdots\\
0\\
-\sqrt{\frac{1}{2}-a}e^{i\gamma_{0}^{1}}
\end{pmatrix},\hspace{0.05cm}|u_{0}^{2}\rangle=\begin{pmatrix}\sqrt{\frac{1}{2}-a}e^{i\alpha_{0}^{2}}\\
0\\
\vdots\\
0\\
-\sqrt{a}e^{i\gamma_{0}^{2}}\\
\sqrt{\frac{1}{2}-a}e^{i\alpha_{0}^{2}}\\
0\\
\vdots\\
0\\
\sqrt{a}e^{i\gamma_{0}^{2}}
\end{pmatrix},\label{zeromodeeigenvectors}
\end{eqnarray}
where only the elements 1,$L+2$, $L+3$, and $2L+4$ are non-null,
$0<a<\frac{1}{2}$ and 
\begin{eqnarray}
\ \alpha_{0}^{1}-\alpha_{0}^{2}=\gamma_{0}^{1}-\gamma_{0}^{2}.\label{angles-1}
\end{eqnarray}
Note that these
eigenstates are independent of the parameters of the $XY$ model and we would like to choose them
in such a way that the Eq. (\ref{Ud}) is preserved. For analytical calculations it should be easier to take all the angles
equal to zero and $a=\frac{1}{4}$. For this choice, we have 
\begin{eqnarray}
\ \eta_{0}=\frac{1}{2}(c_{0}+c_{L+1}+c_{0}^{\dagger}-c_{L+1}^{\dagger}),\label{zeromodeEq1}\\
\eta_{0}^{\dagger}=\frac{1}{2}(c_{0}-c_{L+1}+c_{0}^{\dagger}+c_{L+1}^{\dagger}).\label{zeromodeEq2}
\end{eqnarray}
Depending on the parameters there may be more than one zero mode.
It is important that the eigenstates associated with these zero modes
also preserve the form of the Eq. (\ref{Ud}). Some important examples will appear later in our model. 

It is worth mentioning that if we numerically diagonalize
the matrix $\textbf{M}$ and order the eigenstates according to Eq.
(\ref{lambda}) we have no guarantee that the matrix $\textbf{U}$
will be in the desired form of Eq. (\ref{Ud}). This is because each
eigenstate can be defined with an arbitrary phase, $\exp(i\delta_{l})$.
Thus, we would have to numerically determine the phases to get the matrix
$\textbf{U}$ in the desired form. A simple procedure to obtain the
 matrix $\textbf{U}$ as depicted in the Eq. (\ref{Ud}) is just to
select the first $L+2$ eigenstates associated with the eigenvalues
$\lambda_{i},$ $i=0,1,\ldots,L+1$ and build the matrices $\boldsymbol{g}$
and $\boldsymbol{h}$  and then automatically
we will have the matrix $\textbf{U}$. 

The vacuum state $|\tilde{0}\rangle$ of the Hamiltonian $\textbf{H}_{ff}^{long}$ is now  a state with the following property:
\begin{eqnarray}\label{vaccum}\
\eta_k|\tilde{0}\rangle=0,
\end{eqnarray} 
for all the values of $k$. We also define $N_k=\eta_{k}^{\dagger}\eta_{k}$. Of course because of the zero mode the ground state is degenerate. Therefore one can define the two ground states as
\begin{eqnarray}\label{ground state}\
|\tilde{G}_{\pm}\rangle=\frac{1}{\sqrt{2}}(|\tilde{0}\rangle\pm\eta_0^{\dagger}|\tilde{0}\rangle).
\end{eqnarray} 
%
We will now  show that  $|\tilde{G}_{\pm}\rangle$ are eigenstates of  $\sigma_0^x$  and $\sigma_{L+1}^x$. This fact is important, since we need the eigenstates in the $(+,+)$ sector.

It is easy to show that 
\begin{eqnarray}\label{sigma0 eigenvalue}\
\sigma_0^x|\tilde{G}_{\pm}\rangle=\pm|\tilde{G}_{\pm}\rangle,
\end{eqnarray} 
due to the fact that
\begin{eqnarray}\label{sigma0 }\
\sigma_0^x=c_0+c_0^{\dagger}=\eta_0+\eta_0^{\dagger}.
\end{eqnarray} 
On the other hand, to prove that $|\tilde{G}_{\pm}\rangle$ is  an eigenstate of $\sigma_{L+1}^x$ is not so simple and we need some extra identities, which are presented below.
 
First, note that the following commutation relations hold,
\begin{eqnarray}\label{commutation relations}\
[\sigma_0^x,\sigma_{L+1}^x]=[N_k,\sigma_0^x]=[N_k,\sigma_{L+1}^x]=0,\hspace{0.25cm}k\neq0.
\end{eqnarray} 
To show the last equality we used the fact that for $k\neq0$ we have
\begin{widetext}
\begin{eqnarray}\label{eta eta dagger}\
\eta_k=\sum_{j=0}^{L+1}g_{kj}c_j+h_{kj}c_j^{\dagger}=\sum_{j=0}^{L}(g_{kj}c_j+h_{kj}c_j^{\dagger})+g_{kL+1}(c_{L+1}+c_{L+1}^{\dagger}),\\
\eta_k^{\dagger}=\sum_{j=0}^{L+1}h^*_{kj}c_j+g^*_{kj}c_j^{\dagger}=\sum_{j=0}^{L}(h^*_{kj}c_j+g^*_{kj}c_j^{\dagger})+g^*_{kL+1}(c_{L+1}+c_{L+1}^{\dagger}).
\end{eqnarray} 
\end{widetext}

The last term in both of the above equations means that $\sigma^{y(z)}_{L+1}$ does not appear in the expansion, due to this fact we also have
\begin{eqnarray}\label{commutation relations for fermion operators}\
[\sigma_{L+1}^x,\eta_k]=[\sigma_{L+1}^x,\eta_k^{\dagger}]=0,\hspace{1cm}k\neq0.
\end{eqnarray} 
Using the above equation it is simple to prove the last equality in Eq. (\ref{commutation relations}). There are a few other useful relations that one can prove with a little bit of calculation such as
\begin{eqnarray}\label{anti-commutation relations}\
\{\sigma_0^x,\eta_k\}=\{\sigma_0^x,\eta_k^{\dagger}\}=0,\hspace{1cm}k\neq0.
\end{eqnarray} 
Using Eq. (\ref{commutation relations}), it is now not difficult to prove that 
\begin{eqnarray}\label{sigmaL1 eigenvalue}\
\sigma_{L+1}^x|\tilde{G}_{\pm}\rangle=\delta_{\pm}|\tilde{G}_{\pm}\rangle,
\end{eqnarray} 
where $\delta_{\pm}^2=1$. To have a complete Hilbert space we need to have $\delta_-=-\delta_+$. 

Now one can make the following argument: Consider $\delta_+=+1$, which means $|\tilde{G}_{+}\rangle$ belongs to $(+,+)$ then all the states
\begin{eqnarray}\label{sigmaL1 eigenvalue1}\
\prod_{j=1}^n\eta_{k_j}^{\dagger}|\tilde{G}_{+}\rangle,\hspace{1cm}n \hspace{0.25cm} \text{is even}
\end{eqnarray} 
 also belong to $(+,+)$. Note that in the above equation $0< k_j< k_{j+1}$ which means that the dimension of the space in the sector $(+,+)$ is $2^L$. On the other hand, when $\delta_+=+1$,  $|\tilde{G}_{-}\rangle$ and its tower belongs to the sector $(-,-)$. In other words,
\begin{eqnarray}\label{sigmaL1 eigenvalue2}\
\prod_{j=1}^n\eta_{k_j}^{\dagger}|\tilde{G}_{-}\rangle,\hspace{1cm}n \hspace{0.25cm} \text{is even}
\end{eqnarray} 
belongs to the sector $(-,-)$.
The other two sectors can be built as follows:
\begin{eqnarray}\label{deltaplus extra secors}\
\prod_{j=1}^n\eta_{k_j}^{\dagger}|\tilde{G}_{+}\rangle,\hspace{1cm}n \hspace{0.25cm} \text{is odd},\hspace{1cm}(-,+),\\
\prod_{j=1}^n\eta_{k_j}^{\dagger}|\tilde{G}_{-}\rangle,\hspace{1cm}n \hspace{0.25cm} \text{is odd},\hspace{1cm}(+,-).
\end{eqnarray} 
 Similarly one can show that if 
 $|\tilde{G}_{+}\rangle$ belongs to $(+,-)$ which means $\delta_+=-1$, then one can write
\begin{eqnarray}\label{delta minus sectors}\
\prod_{j=0}^n\eta_{k_j}^{\dagger}|\tilde{G}_{+}\rangle,\hspace{1cm}n \hspace{0.25cm} \text{is even},\hspace{1cm}(+,-),\\
\prod_{j=0}^n\eta_{k_j}^{\dagger}|\tilde{G}_{-}\rangle,\hspace{1cm}n \hspace{0.25cm} \text{is even},\hspace{1cm}(-,+),\\
\prod_{j=1}^n\eta_{k_j}^{\dagger}|\tilde{G}_{+}\rangle,\hspace{1cm}n \hspace{0.25cm} \text{is odd},\hspace{1cm}(-,-),\\
\prod_{j=1}^n\eta_{k_j}^{\dagger}|\tilde{G}_{-}\rangle,\hspace{1cm}n \hspace{0.25cm} \text{is odd},\hspace{1cm}(+,+).
\end{eqnarray} 

 The above argument means that to know the sector $(+,+)$ we need to figure out the value of $\delta_+$. The ground state of the Hamiltonian $H^{XY}$
 is going to be one of the following two states of the $H^{long}$:
\begin{eqnarray}\label{delta minus sectors 2}\
&|\tilde{G}_{+}\rangle&\hspace{1cm}\delta_+=1,\\
&\eta_{k_{\text{min}}}^{\dagger}|\tilde{G}_{-}\rangle&\hspace{1cm}\delta_+=-1.
\label{delta minus sectors 3}\
\end{eqnarray} 

In principle, to find the right ground state we need to calculate $\delta_+$ as follows:
\begin{eqnarray}\label{delta plus}\
\delta_+=\langle\tilde{G}_{+}|\sigma_{L+1}^x|\tilde{G}_{+}\rangle.
\end{eqnarray} 
The value of $\delta_+$ can be found by a bit of manipulations and using the Wick theorem. The detail of the calculation is presented in the Appendix \ref{SEC:delta-Plus}. 
We observed that the ground state energies of  transverse field Ising chain [$\gamma=1$ and $h=1$ in Eq. (\ref{HXY})] and the XX chain  
[$\gamma=0$ and $h=0$ in Eq. (\ref{HXY})] correspond to the first excited states energies  of the $\textbf{H}_{ff}^{long}$ when the boundary magnetic field on both boundaries are the same and $J=+1$. In other words the ground state of these two models are basically the state $\eta_{k_{\text{min}}}^{\dagger}|\tilde{G}_{-}\rangle$ with $k_{\text{min}}=1$ of $\textbf{H}_{ff}^{long}$.

\subsection{Correlation matrices}
\label{subsec:correlation}
In this section, we calculate the correlation functions that are necessary to calculate the entanglement entropy. In principle we need both $\langle\tilde{G}_{+}|\mathcal{O}|\tilde{G}_{+}\rangle$ and $\langle\tilde{G}_{-}|\eta_{k_{\text{min}}}\mathcal{O}\eta_{k_{\text{min}}}^{\dagger}|\tilde{G}_{-}\rangle$, where $\mathcal{O}$ is the one and two point functions of the fermionic operators. The rest of the correlations can be reproduced with proper use of the Wick's theorem. An easy calculation shows that:
\begin{eqnarray}\label{one point function}
\langle\tilde{G}_{+}|c_{j}|\tilde{G}_{+}\rangle&=&\frac{1}{2}(g_{0j}^{*}+h_{0j}),\\
\langle\tilde{G}_{+}|c_{j}^{\dagger}|\tilde{G}_{+}\rangle&=&\frac{1}{2}(h_{0j}^{*}+g_{0j}).
\end{eqnarray}
Then because of Eqs. (\ref{zeromodeEq1}) and (\ref{zeromodeEq2}) one can write
\begin{eqnarray}\label{one point function2}
\langle\tilde{G}_{+}|c_{j}|\tilde{G}_{+}\rangle & = & \frac{1}{2}\delta_{0,j},\\
\langle\tilde{G}_{+}|c_{j}^{\dagger}|\tilde{G}_{+}\rangle & = &\frac{1}{2}\delta_{0,j}.
\end{eqnarray}
As expected,  due to the fact that the spin at site zero is
in the positive direction of $\sigma^x_{0}$ the above expectation values are zero for $j=1,2,...,L,L+1$.
For the expectation values of $\langle\tilde{G}_{-}|\eta_{k_{\text{min}}}c_{j}(c_{j}^{\dagger})\eta_{k_{\text{min}}}^{\dagger}|\tilde{G}_{-}\rangle$ the same result is correct, as it is expected.  
 
To proceed and calculate the two point correlation functions, we first define the 
$\boldsymbol{\Gamma}$ matrix as a block matrix which is built from the correlation functions as follows
\begin{eqnarray}\label{gamma1}\
\boldsymbol{\Gamma}_{ln}=\begin{pmatrix}
\langle a_{l}^{x}a_{n}^{x} \rangle -I_{l\times n} & \langle a_{l}^{x}a_{n}^{y} \rangle \\
\langle a_{l}^{y}a_{n}^{x} \rangle  &\langle a_{l}^{y}a_{n}^{y} \rangle -I_{l\times n} \\
  \end{pmatrix},\hspace{0.55cm}
\end{eqnarray} 
where $a_{l}^{x}=c_{l}^{\dagger}+c_{l}$ and $a_{l}^{y}=i(c_{l}-c_{l}^{\dagger})$. One can easily find all the different elements of the $\boldsymbol{\Gamma}$ matrix.
It is convenient to write the $\boldsymbol{\Gamma}$ matrix, as

\begin{eqnarray}\label{gamma11}\
\boldsymbol{\Gamma}=\begin{pmatrix}
\boldsymbol{\Gamma}^{11} & \boldsymbol{\Gamma}^{12} \\
\boldsymbol{\Gamma}^{21}  &\boldsymbol{\Gamma}^{22} \\
  \end{pmatrix},\hspace{0.55cm}
\end{eqnarray} 
where the matrices $\boldsymbol{\Gamma}^{ij}$ of dimension
$(\ell+2)\times(\ell+2)$, $\ell=0,1,...,L+2$, are
\begin{eqnarray}\label{gamma3}
\boldsymbol{\Gamma}^{11}&=&\boldsymbol{F}+\boldsymbol{F}^{\dagger}+\boldsymbol{C}-\boldsymbol{C}^{T},\\
\boldsymbol{\Gamma}^{12}&=&i(-I+\boldsymbol{C}+\boldsymbol{C}^{T}-\boldsymbol{F}+\boldsymbol{F}^{\dagger}), \\
\boldsymbol{\Gamma}^{21}&=&-i(-I+\boldsymbol{C}+\boldsymbol{C}^{T}+\boldsymbol{F}-\boldsymbol{F}^{\dagger}), \\
\boldsymbol{\Gamma}^{22}&=& -\boldsymbol{F}-\boldsymbol{F}^{\dagger}+\boldsymbol{C}-\boldsymbol{C}^{T},
\end{eqnarray}
where $F_{ln}=\langle c_{l}^{\dagger}c_{n}^{\dagger}\rangle$ and $C_{ln}=\langle c_{l}^{\dagger}c_{n}\rangle$.
For the states $|\tilde{G}_{\pm}\rangle$ we have 
\begin{eqnarray}\label{fln}
F_{ln}^{\pm}&=&\langle \tilde{G}_{\pm} | c_{l}^{\dagger}c_{n}^{\dagger}|\tilde{G}_{\pm} \rangle=(h^{\dagger}g)_{ln}+\frac{1}{2}(g_{0,l}h^*_{0,n}-h^*_{0,l}g_{0,n}),\hspace{0.58cm}\\
C_{ln}^{\pm}&=&\langle \tilde{G}_{\pm} |c_{l}^{\dagger}c_{n}|\tilde{G}_{\pm}\rangle=(h^{\dagger}h)_{ln}+\frac{1}{2}(g_{0,l}g^*_{0,n}-h^*_{0,l}h_{0,n}).\hspace{0.58cm}
\end{eqnarray}

While for the  states $\eta_{k_{\text{min}}}^{\dagger}|\tilde{G}_{\pm}\rangle$ we have
\begin{widetext}
\begin{eqnarray}\label{fln-min}
F^{ex\pm}_{ln}&=&\langle \tilde{G}_{\pm} |\eta_{k_{\text{min}}} c_{l}^{\dagger}c_{n}^{\dagger}\eta_{k_{\text{min}}}^{\dagger}|\tilde{G}_{\pm} \rangle=(h^{\dagger}g)_{ln}+\frac{1}{2}\sum_{j=0,1}(g_{j,l}h^*_{j,n}-h^*_{j,l}g_{j,n})(j+1),\\
C^{ex\pm}_{ln}&=&\langle \tilde{G}_{\pm} |\eta_{k_{\text{min}}}c_{l}c_{n}^{\dagger}\eta_{k_{\text{min}}}^{\dagger}|\tilde{G}_{\pm}\rangle=(h^{\dagger}h)_{ln}+\frac{1}{2}\sum_{j=0,1}(g_{j,l}g^*_{j,n}-h^*_{j,l}h_{j,n}) (j+1).
\end{eqnarray}
\end{widetext}
Note that $F^{-}_{ln}=F^{+}_{ln}$ and  $C^{-}_{ln}=C^{+}_{ln}$ as well as $F^{ex-}_{ln}=F^{ex+}_{ln}$ and  $C^{ex-}_{ln}=C^{ex+}_{ln}$. Due to these results, it is expected that both of the two degenerate sectors give the same results for the entanglement entropy.

With a little bit of calculation one can also show that as far as $\mathcal{O}$ does not have $c_{0}$ and
$c_{0}^{\dagger}$ then we have
\begin{eqnarray}\label{Wick theorem general}
\langle \tilde{G}_{+} |\mathcal{O}|\tilde{G}_{+}\rangle&=&\langle \tilde{0} |\mathcal{O}|\tilde{0}\rangle,\\
\langle \tilde{G}_{+} |c_0^{\dagger}c_0\mathcal{O}|\tilde{G}_{+}\rangle&=&\langle \tilde{G}_{+} |c_0\mathcal{O}|\tilde{G}_{+}\rangle.
\end{eqnarray}
If $\mathcal{O}$ is made of multiplication of even number of creation and annihilation operators we have
\begin{eqnarray}\label{Wick theorem general even}
\langle \tilde{G}_{+} |c_0\mathcal{O}|\tilde{G}_{+}\rangle&=&\frac{1}{2}\langle \tilde{0} |\mathcal{O}|\tilde{0}\rangle,\\
\langle \tilde{G}_{+} |c_0^{\dagger}\mathcal{O}|\tilde{G}_{+}\rangle&=&\frac{1}{2}\langle \tilde{0} |\mathcal{O}|\tilde{0}\rangle,\\
\langle \tilde{G}_{+} |c_0^{\dagger}c_0\mathcal{O}|\tilde{G}_{+}\rangle&=&\frac{1}{2}\langle \tilde{0} |\mathcal{O}|\tilde{0}\rangle.
\end{eqnarray}
On the other hand, if $\mathcal{O}$ is made of multiplication of odd number of creation and annihilation operators we have
\begin{eqnarray}\label{Wick theorem general even 2}
\langle \tilde{G}_{+} |c_0\mathcal{O}|\tilde{G}_{+}\rangle&=&\langle \tilde{0} |c_0\mathcal{O}|\tilde{0}\rangle,\\
\langle \tilde{G}_{+} |c_0^{\dagger}\mathcal{O}|\tilde{G}_{+}\rangle&=&\langle \tilde{0} |c_0^{\dagger}\mathcal{O}|\tilde{0}\rangle.
\end{eqnarray}
The right hand side of the above equations can be calculated easily by using directly the Wick's theorem.
Finally it is easy to see that the elements of the first row and column of the $\boldsymbol{\Gamma}$ matrix are all zero if we include the site zero.

\subsection{Entanglement entropy}
\label{subsec:entanglement}

In this section, we explain how one can use the $\boldsymbol{\Gamma}$ matrix to calculate the entanglement entropy of a subsystem  that starts from one boundary. We emphasize that to calculate the entanglement entropy of the XY chain with ADBMF,  we had to generalize the Peschel method \cite{Chung2001,peschel2003calculation,Vidal2001,Jin2004}.
We need to make a small adjustment to the Peschel method  because after the projection the two ghost sites are not entangled with the rest of the system and we can not write the projected state in an exponential form.
 In addition the odd point functions of the
fermionic operators with the site zero included is non-zero too.


The main idea of the Peschel method is to connect the entanglement
entropy to the eigenvalues of the $\boldsymbol{\Gamma}$ matrix and
exploit the Wick's theorem. Here, we need to take into account the fact that the odd point functions are non-zero for the site zero.  Since the $\boldsymbol{\Gamma}$
matrix is a skew symmetric matrix it can be block diagonalized using
an orthogonal matrix $\textbf{V}$ as 

\begin{eqnarray}
\ \textbf{V}\boldsymbol{\Gamma}\textbf{V}^{T}=
\begin{pmatrix}0 & i\boldsymbol{\nu}\\
-i\boldsymbol{\nu} & 0
\end{pmatrix}.\label{block diagonal gamma}
\end{eqnarray}
where $\boldsymbol{\nu}$ is a diagonal matrix. Then we can define
the following fermionic operators: 
\begin{eqnarray}
\ \begin{pmatrix}\textbf{d}^{\dagger}\\
\textbf{d}
\end{pmatrix}=\frac{1}{2}\begin{pmatrix}I & iI\\
I & -iI
\end{pmatrix}\textbf{V}\begin{pmatrix}\textbf{a}^{x}\\
\textbf{a}^{y}
\end{pmatrix},\label{d fermions}
\end{eqnarray}
with the correlation matrices 
\begin{eqnarray}
\ \Big{\langle}\begin{pmatrix}\textbf{d}^{\dagger}\\
\textbf{d}
\end{pmatrix}\begin{pmatrix}\textbf{d} & \textbf{d}^{\dagger}\end{pmatrix}\Big{\rangle}=\begin{pmatrix}\frac{I+\boldsymbol{\nu}}{2} & 0\\
0 & \frac{I-\boldsymbol{\nu}}{2}
\end{pmatrix}.\label{correlation matrices d}
\end{eqnarray}
Using the equation (\ref{d fermions}) one can also express the one point
function of fermionic operators $d_k$ and $d_k^{\dagger}$   in terms  of the elements
of the matrices $\textbf{g}$, $\textbf{h}$ and $\textbf{V}$. 
%
Numerical investigation shows that
\begin{eqnarray}\label{one-point-d}
\langle d_{k}\rangle&=&\frac{1}{2} T_{k,0}\delta_{k,0},\\
\langle d_{k}^{\dagger}\rangle&=&\frac{1}{2}T_{k,0}^*\delta_{k,0},
\end{eqnarray}
where the matrix \textbf{T} is related with the unitary transformation \textbf{W}, which diagonalizes $\boldsymbol{\Gamma}$, by
\begin{eqnarray}\label{T matrix}
\textbf{T} =2\textbf{W}
\begin{pmatrix}
I & iI\\
I & -iI
\end{pmatrix}. 
\end{eqnarray}\label{one-point-d2}
%

Note that we also have $\nu_{0}=0$. Having the above results in
hand one can make the following ansatz for the reduced density matrix
of the subsystem $A$ with size $\ell$ $(\ell=0,1,2,..)$: 
\begin{eqnarray}
\ \rho_{A}(\ell)=\frac{T_{00}^*d_{0}+T_{00}d_{0}^{\dagger}+I}{2}\times\hspace{2cm}\nonumber\\\prod_{k=1}^{\ell}\Big{(}\frac{1+\nu_{k}}{2}d_{k}^{\dagger}d_{k}+\frac{1-\nu_{k}}{2}d_{k}d_{k}^{\dagger}\Big{)}.\hspace{0.25cm
}\label{reduced density matrix}
\end{eqnarray}
 We note that the above ansatz also respects the generalized Wick's theorem that we introduced in the previous section, which
means that this reduced density matrix produces all the correlation
functions correctly.  Note that the reduced
density matrix $\rho_{A}(0)$  is built with the eigenstate of $\sigma_{x}$ associated with the eigenvalue $+1$. This is evident by looking at the identity 
\begin{eqnarray}
T_{00}^*d_{0}+T_{00}d_{0}^{\dagger}=c_{0}+c_{0}^{\dagger}
\end{eqnarray}
which we confirmed numerically for various values of the parameters.

Finally, the entanglement entropy
can be written as: 
\begin{eqnarray}
\ S=-\sum_{k=1}^{\ell}[\frac{1+\nu_{k}}{2}\ln\frac{1+\nu_{k}}{2}+\frac{1-\nu_{k}}{2}\ln\frac{1-\nu_{k}}{2}],\label{entanglement entropy}
\end{eqnarray}
which can be also recast as: 
\begin{equation}
S=-{\rm Tr}[\frac{1+\boldsymbol{\Gamma}}{2}\ln{(\frac{1+\boldsymbol{\Gamma}}{2}})]-\ln2.\label{ent2-1}
\end{equation}
In the appendix \ref{SEC:ln2}, we present an argument showing that if the reduced density matrix is build from a pure state like $\mid\Psi_{k}^{L}>=\mid+>\otimes\mid\Psi_{k}^{L-1}>$ we  need to subtract  a  $\ln2$ term in the entanglement entropy in order to get the right result. It is convenient to mention that a similar $\log 2$ subtraction in the
entanglement entropy was also observed in the context of quantum quench of the
XY chain \cite{PhysRevB.98.161117}. 
For periodic and open systems that one does not need to do any projection in the state the standard method, which does not need the subtraction of $\ln2$, works as usual \cite{Chung2001,peschel2003calculation,Vidal2001,Jin2004}.

\subsection{Boundary entropy: Transverse field Ising chain}
\label{subsec:XYBE}
In this subsection, we present our numerical estimates of the ground state
degeneracy $g\equiv g(\theta,\phi)$ of the critical Transferse field Ising
chain when we have arbitrary equal boundary
fields on the two edges. 
Although the method presented in the last section works for arbitrary BCs, in this subsection we consider the transverse field Ising chain with equal boundary conditions on the two edges.

\begin{figure}
\includegraphics[scale=0.45]{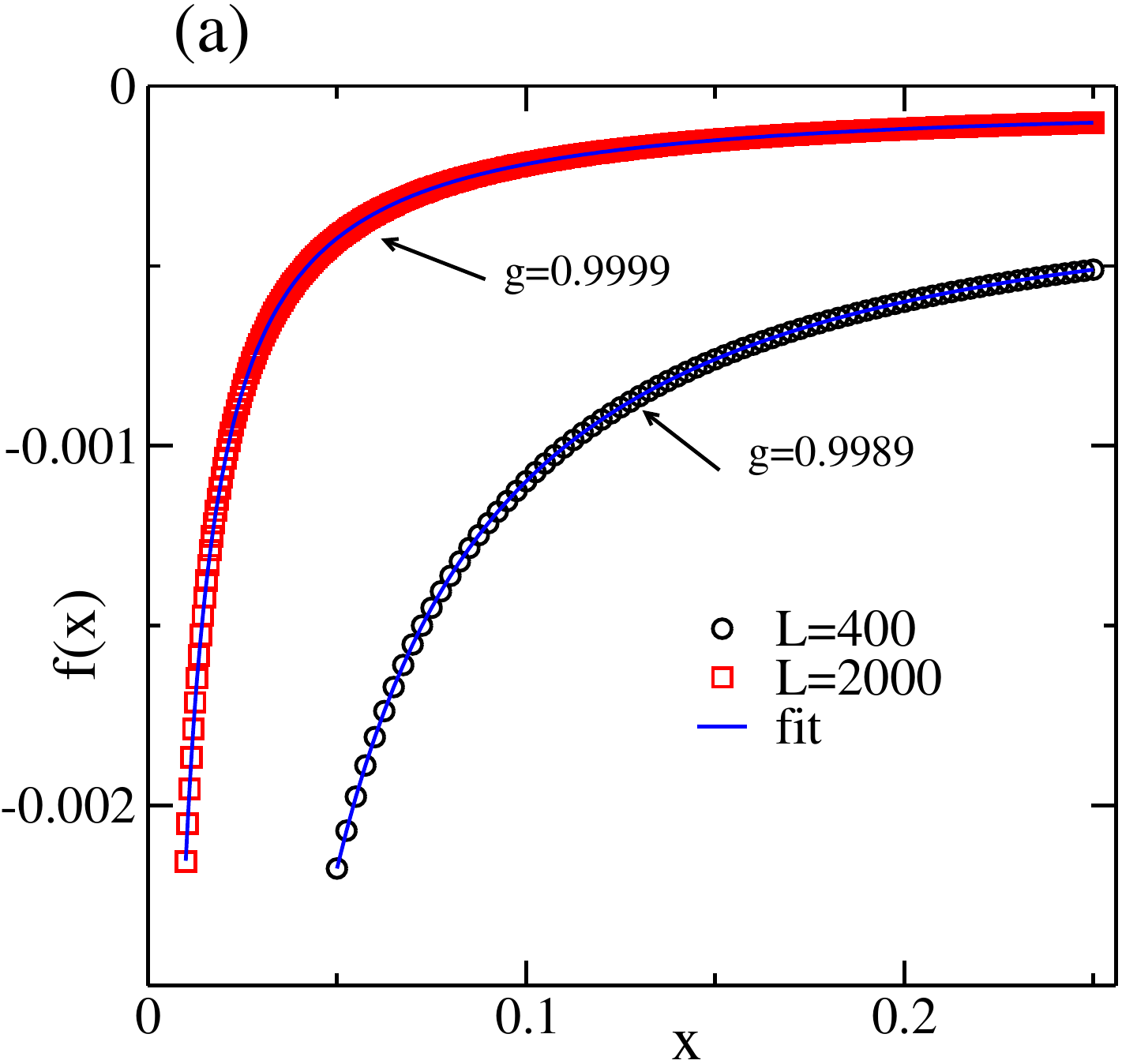} \includegraphics[scale=0.45]{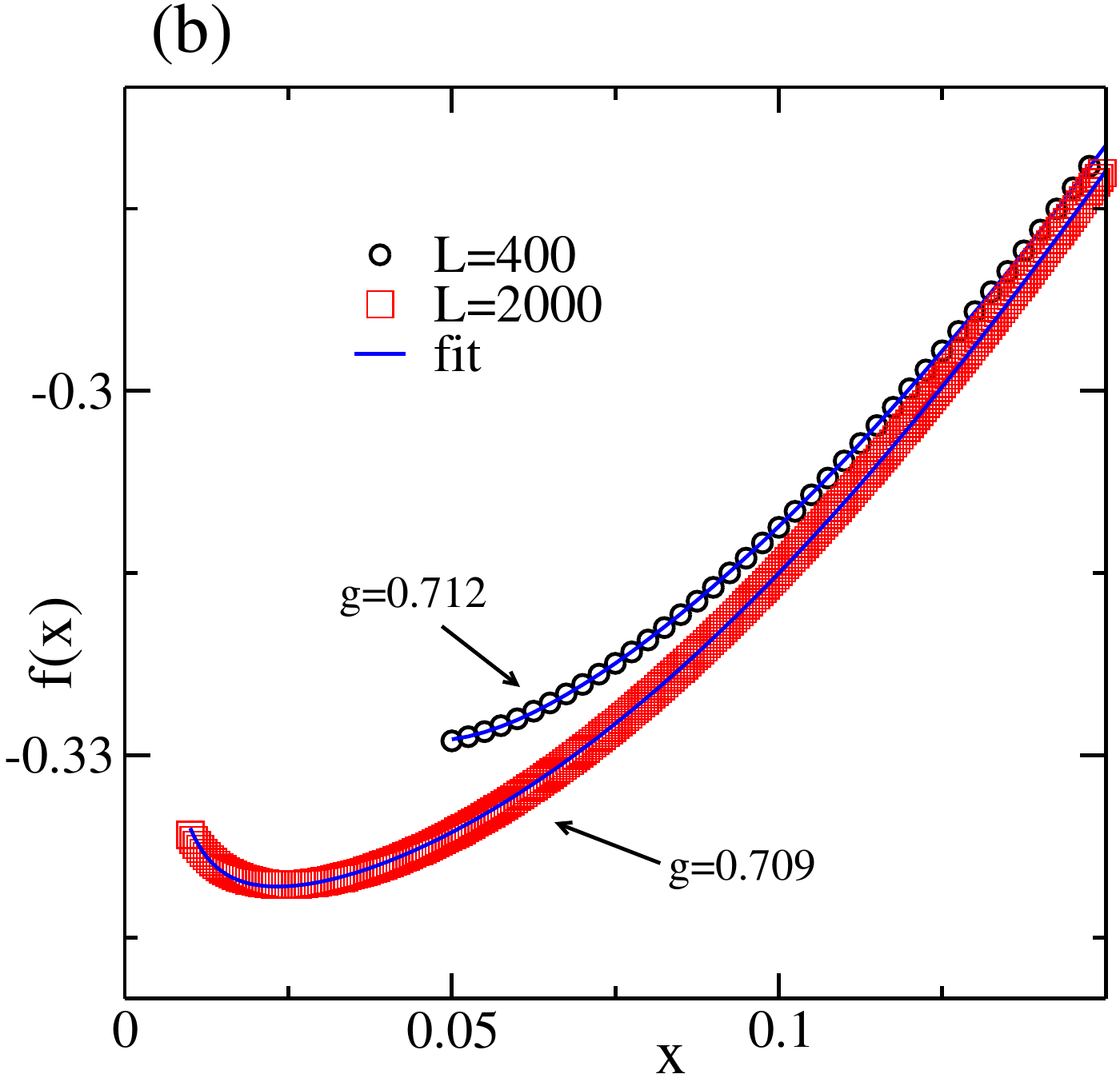} 

\caption{\label{fig1Ising}The function $f(x)$ vs. $x$ for the transverse field Ising
chain and two system sizes (see legends). The symbols are the numerical
data and the continuous lines are the best fit to Eq. (\ref{fx}).
The arrows indicate the values of $g$ we get by the fitting procedure.
(a) Results for the OBC case with zero boundary magnetic fields. (b) Results for the boundary magnetic
field case with $\theta=\pi/4$, $\phi=\pi/4$ and $b=4$.}
\end{figure}

In order to show that we are able to find quite good estimates
of the ground state degeneracy $g$ by fitting the numerical data
to Eq. (\ref{fx}), we first consider the transverse field Ising chain with
OBC {[}Eq. (\ref{HXY}) with $\gamma=1$, $\vec{b}_{1}=\vec{b}_{L}=0$
and $J=+1$] whose exact value of the ground state degeneracy is $g_{free}^{Ising}=1$
\citep{CARDY1989,CARDY1991,Barthel2006}. 
It is worth  mentioning for $J<0$ we have $G_b=0$ when the edge magnetic fields are the same and are in the $x$ direction, see Ref. \cite{Taddia2013}. However, for positive values of $J$ we observed that the non-universal function $G_b$ is non-zero. 
%
In Fig. \ref{fig1Ising}(a), we present the function $f(x)$ that we calculated numerically by using the
correlation matrix method, explained in the previous section, for $L=400$ and $L=2000$. The anticipated scaling behaviors of the entanglement entropies in Eqs. (\ref{EEpbc}) and (\ref{EEb}) hold for $\ell\gg 1$. Moreover, Eq. (\ref{sb}) is valid if $x=\ell/L\ll 1$.
Due to these reasons, we fit the numerical data considering $20<\ell<0.25L$. As we can see in Fig. \ref{fig1Ising}(a), we are able to fit quite well the numerical data with Eq. (\ref{fx}) and the obtained estimates of $g$ are very close to the expected exact one, i. e. $g=1$.

Now, we discuss the case of arbitrary direction of the boundary magnetic
fields. 
As mentioned before, we are going to consider that the boundary magnetic fields
are the same in both edges, i. e., $b_{1}=b_{L}=b$, $\theta_{1}=\theta_{L}=\theta$
and $\phi_{1}=\phi_{L}=\phi$. Then, the magnitude of the effective
boundary magnetic fields in the directions $x$, $y$ and $z$ are $h_{x}^{b}=b\sin\theta\cos\phi$,
$h_{y}^{b}=b\sin\theta\sin\phi$, and $h_{z}^{b}=b\cos\theta$, respectively
for both edges. For systems with boundaries we must be careful when we use the Eq. (\ref{fx})  to get estimates of $g$, since Eq. (\ref{EEb})  holds only if the crossover lengths $\xi\sim h_b^{d-1}<\ell$.  The scaling dimension  of the relevant boundary perturbation (in the $x$ direction) for the transverse field Ising chain is $d_x=1/2$ \cite{CARDY1989,CARDY1991}. For instance, we need to be careful when we estimate $g$
for $\theta\rightarrow0$ and/or $\phi\rightarrow\pi/2$, since the
crossover length $\xi^{x}\thicksim(b\sin\theta\cos\phi)^{-1/2}$,
for a \emph{finite boundary magnetic field}, can diverge in these regimes. Due to this
facts, we expect a huge crossover effect mainly if $\theta\rightarrow0$
\emph{and/or} $\phi\rightarrow\pi/2$ since $\xi^{x}\thicksim1/\sqrt{b\theta\left(\pi/2-\phi\right)}$. 
In most of the cases, we observed that for the interval $1\leq b\leq10$, the estimate of $g$ changes  very little, which is indicative that we are in the regime that $\xi<\ell$.

It is also important to mention that for 
$\phi=\pi/2$ and $\theta\in$$(0,\pi/2]$ the matrix $M$ has
four zero eigenvalues, while for $\theta=0$ and $\phi\in$$[0,\pi/2]$
has six zero eigenvalues. In these two regimes, the correlation matrix  approach, presented before, to obtain
the entanglement entropy needs a bit of modification, because if one does not take into account the extra degeneracies  the matrix $U$ will
not necessarily be in the desired canonical form, see Eq. (\ref{Ud}).
We will analyze those situations later. For other values of $\theta$
and $\phi$ we found that the matrix $M$ has just two zero eigenvalues,
whose eigenstates are given in the Eq. (\ref{zeromodeeigenvectors}).

\begin{figure}
\includegraphics[scale=0.45]{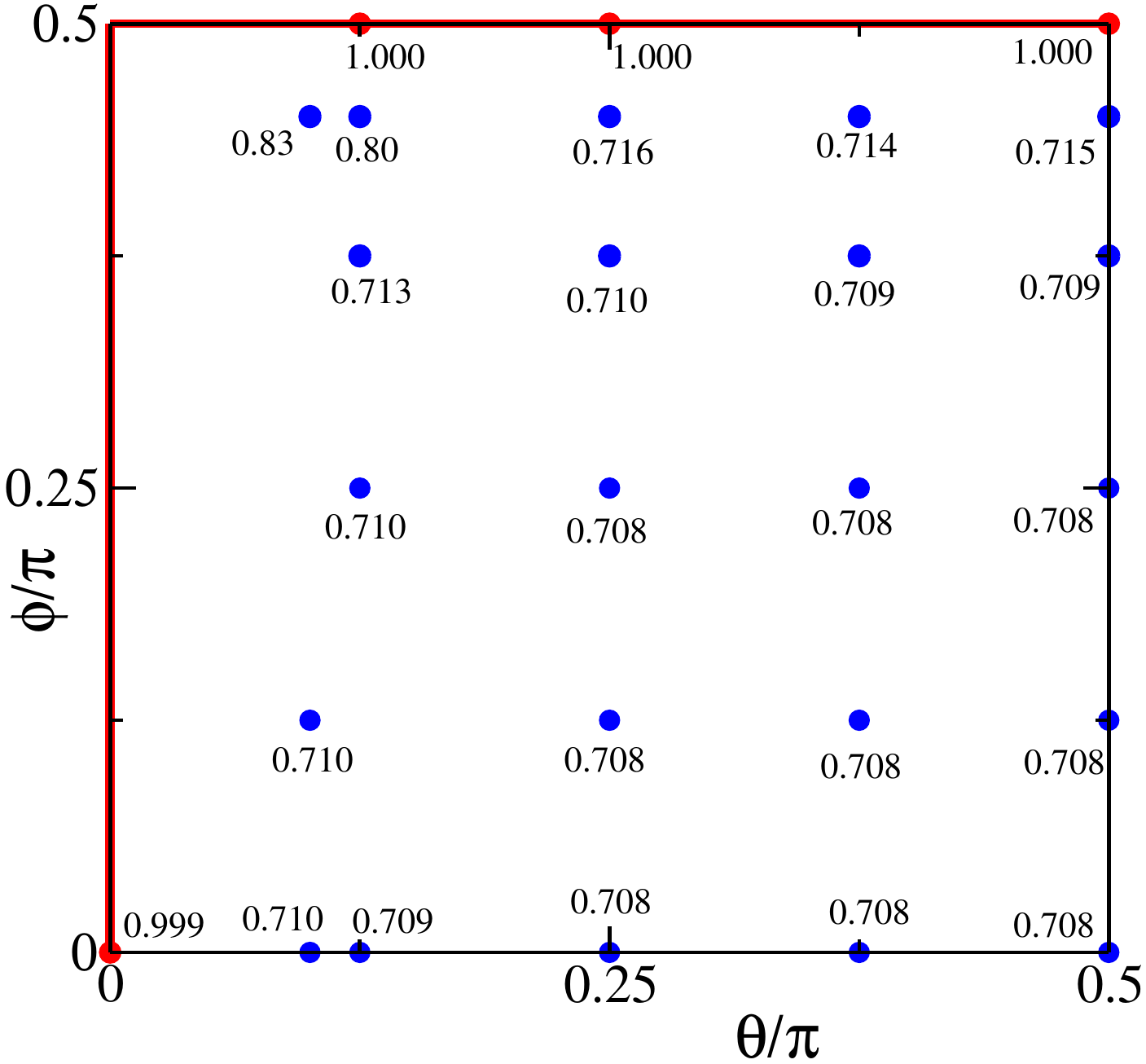} 

\caption{\label{figg} The numerical estimates of the ground state degeneracy $g(\theta,\phi)$
of the transverse field Ising chain for $L=2000$ and $b=4$. For $\phi=0.45$
and $\theta=0.1\pi$ and $\theta=0.125\pi$ exceptionally we used $b=500$. The circles are
the points $(\theta,\phi)$ that we consider and the values close to them
are the estimates of $g$ we get by the fitting procedure (see text).
The results indicate that along the red lines we have $g=1$ and away from
these lines we have $g=\sqrt{2}/2=0.707.$ }
\end{figure}

In Figs. \ref{fig1Ising}(b), we present a representative result of the
function $f(x)$ for the ADBFM case. For this particular example, where $\phi=\pi/4$, $\theta=\pi/4$
and $b=4$, we get $g\sim0.709$ for $L=2000$, which is very close
to the expected exact value $g_{fixed}^{Ising}=\sqrt{2}/2=0.7071...$ \cite{CARDY1989,CARDY1991,Barthel2006}.
Using the explained fitting procedure, we estimated $g(\theta,\phi)$ for
several other values of angles for system sizes $L=2000$. The obtained values
are depicted in Fig. \ref{figg}. As we can see in this figure, for
$\phi\neq\pi/2$ and $\theta\notin$$(0,\pi/2]$ as well as for $\theta\neq0$
and $\phi\notin$$[0,\pi/2]$ the results strongly indicate that $g(\theta,\phi)=\sqrt{2}/2$,
except for small values of $\theta$ and $\phi$ close to $\pi/2$.
As we already mentioned, in this region we expected a huge crossover
effect for a finite boundary magnetic field. For instance, for $\theta=0.1\pi$
and $\phi=0.45\pi$, and considering $b=500$ we got $g=0.94$ and
$g=0.83$ for $L=400$ and $L=2000$, respectively. We also observed
that for a fixed value of $L$ the estimate of $g$ depends on the
value of $b$. These finite-size effects are indicative that even
for $b=500$, and $L=2000$ we are still not in the regime that $\xi^{x}<\ell$ for this angles. 

Now, we consider the case $\theta=0$, where we have $b_{i}\cdot\vec{S}_{i}=\frac{b_{1}}{2}\sigma_{i}^{z}=b_{i}\left(c_{i}^{\dagger}c_{j}-1/2\right)$,
$i=1$ and $L$. In this case, $b_{i}\cdot\vec{S}_{i}$ is quadratic
in terms of creation and annihilations operators, so it is possible
to map $H^{XY}$ to a quadratic free fermion Hamiltonian, and we can use the standard matrix correlation method to obtain the entanglement entropy.
Since this perturbation is not relevant,
we expect that the BE in this case be the same as the OBC case, i. e., $s_{b}=0$ (or equivalently $g=1).$ Indeed, our numerical estimates of $g$, based on the fitting procedure, agree  very well with the expected value. For $b=4$ we get $g=0.998$ and
$g=0.999$ for $L=400$ and $L=2000,$ respectively.

Finally, we discuss the case $\phi=\pi/2$. As we mentioned before,
in this case we have four zero eigenvalues. Two eigenvectors, associated with these eigenvalues, are those
 given in Eq. (\ref{zeromodeeigenvectors}) and the other two
are 
\begin{eqnarray}
\ |u_{0}^{1}\rangle=\frac{1}{N_{nor}}\begin{pmatrix}\alpha i\\
2\\
0\\
\vdots\\
0\\
-2\\
-\alpha i\\
-\alpha i\\
2\\
0\\
\vdots\\
0\\
2\\
-\alpha i
\end{pmatrix},\hspace{0.5cm}|u_{0}^{2}\rangle=\frac{1}{N_{nor}}\begin{pmatrix}\alpha i\\
2\\
0\\
\vdots\\
0\\
2\\
\alpha i\\
-\alpha i\\
2\\
0\\
\vdots\\
0\\
-2\\
\alpha i
\end{pmatrix},\label{zeromodeeigenvectors-2}
\end{eqnarray}
where $\alpha=2b\left(\cot\theta-\sec\theta\right)$ and $N_{nor}=2\sqrt{2\alpha^{2}+4}.$
The boundary perturbation now is given by $b_{i}\cdot\vec{S}_{i}=\frac{b_{1}}{2}\left(\sin\theta\sigma_{i}^{y}+\cos\theta\sigma_{i}^{z}\right),$
$i=1$ and $L$, and is not relevant too. Due to this fact, here too we
expect that $g=1$ along the line with $\phi=\pi/2$. Again, our
numerical data supports this prediction. We found that $g\sim1.000$  along
this line. 

In summary as far as the boundary magnetic field vector is in the $yz$ plane one gets free boundary condition. Introducing even a small boundary  magnetic field in the $x$ direction which means breaking the bulk $Z_2$ symmetry induces a fixed boundary condition.





\section{The XXZ chain with arbitrary direction of the boundary magnetic field}
\label{sec:XXZ}

In this section, we investigate the spin-1/2 $XXZ$ chain with ADBMF
given by


\begin{eqnarray}
\ \textbf{H}^{XXZ}=J\sum_{j=1}^{L-1}\Big{[}S_{j}^{x}S_{j+1}^{x}+S_{j}^{y}S_{j+1}^{y}+\Delta S_{j}^{z}S_{j+1}^{z}\Big{]}\nonumber\\
+\vec{b}_{1}\cdot\vec{S}_{1}+\vec{b}_{L}\cdot\vec{S}_{L}\,,\hspace{1.8cm}\label{HXYZ}
\end{eqnarray}
where $\Delta$ is the anisotropy and we
use $J=1$ in order to fix the energy scale. The boundary magnetic
fields $\vec{b}_{i}$, $i=1$ and $L$, are defined in Eqs. (\ref{b1}) and
(\ref{b2}) and we consider that the magnitude of both boundary  magnetic fields are the same, i. e. $b=b_1=b_L$. For $-1<\Delta\le1$ the system is bulk critical with central
charge $c=1$. The OBC case corresponds to the free conformally invariant boundary condition with $g=g^{OBC}=\frac{1}{\pi^{1/4}\sqrt{2R}}$, where $R^2=\frac{1}{2\pi}\left(1-\frac{\arccos\Delta}{\pi}\right)$ \cite{affleckboundaryxxz}.
While the fixed conformally invariant
boundary condition with $g=g^{fixed}=\pi^{1/4}\sqrt{R}$ corresponds to the case that both boundary magnetic fields are in the $x$
direction ($\phi_{1}=\phi_{2}=0$ and $\theta=\frac{\pi}{2}$) with
$b_{1}=b_{L}=\infty$ \cite{affleckboundaryxxz}.
Note that these predictions were obtained by bosonization technique and to our knowledge they were not verified by other entanglement approaches. 
It is important to mention that although the XXZ chain with ADBMF is exactly solvable by the thermodynamic Bethe ansatz method\cite{Vega1993,NEPOMECHIE2002615,Nepomechie2003,CAO2003487,CAO2013152,CAO20131522,NICCOLI2013397,Faldella2014,Belliard2013,Kitanine2014,Nepomechie2013-2,LI201417,Pozsgay2018}, it seems in the massless regime the determination of the  free boundary energy in the low temperature regime is not simple\footnote{We thanks B. Pozsgay for pointing this fact to us.}.
Thus, it is highly  desirable to confirm the bosonization prediction by other unbiased techniques. 
Furthermore, there is no prediction of the values of $g$ for other directions of the boundary magnetic fields. We intent to provide further insight about these issues, in this section. 
It is also important to mention that, like the Ising case when the edge magnetic fields are the same and  are in the $x$ direction, here too if we choose $J<0$ the corrections in $S^b(L,\ell)$  are zero, i. e. $G_b=0$. However, for positive values of $J$ those corrections are present.    
 
Before starting to present the results, it is convenient to mention that in the case that both boundary magnetic fields are the same, we verified, numerically, that the energies as well as the entanglement entropy of the XXZ chain do not depend on the values of the angle $\phi=\phi_{1}=\phi_{L}$. Another important point is that in the case of the  XX chain with ADBMF {[}$\Delta=0$ in Eq. (\ref{HXYZ}){]}, the Hamiltonian is the same  as the one in Eq. (\ref{HXY})
with $\gamma=0$ and $h=0$. Consequently, we can use the correlation matrix
method developed in the previous section to obtain the entanglement entropy.

\begin{figure}[!h]
\includegraphics[scale=0.45]{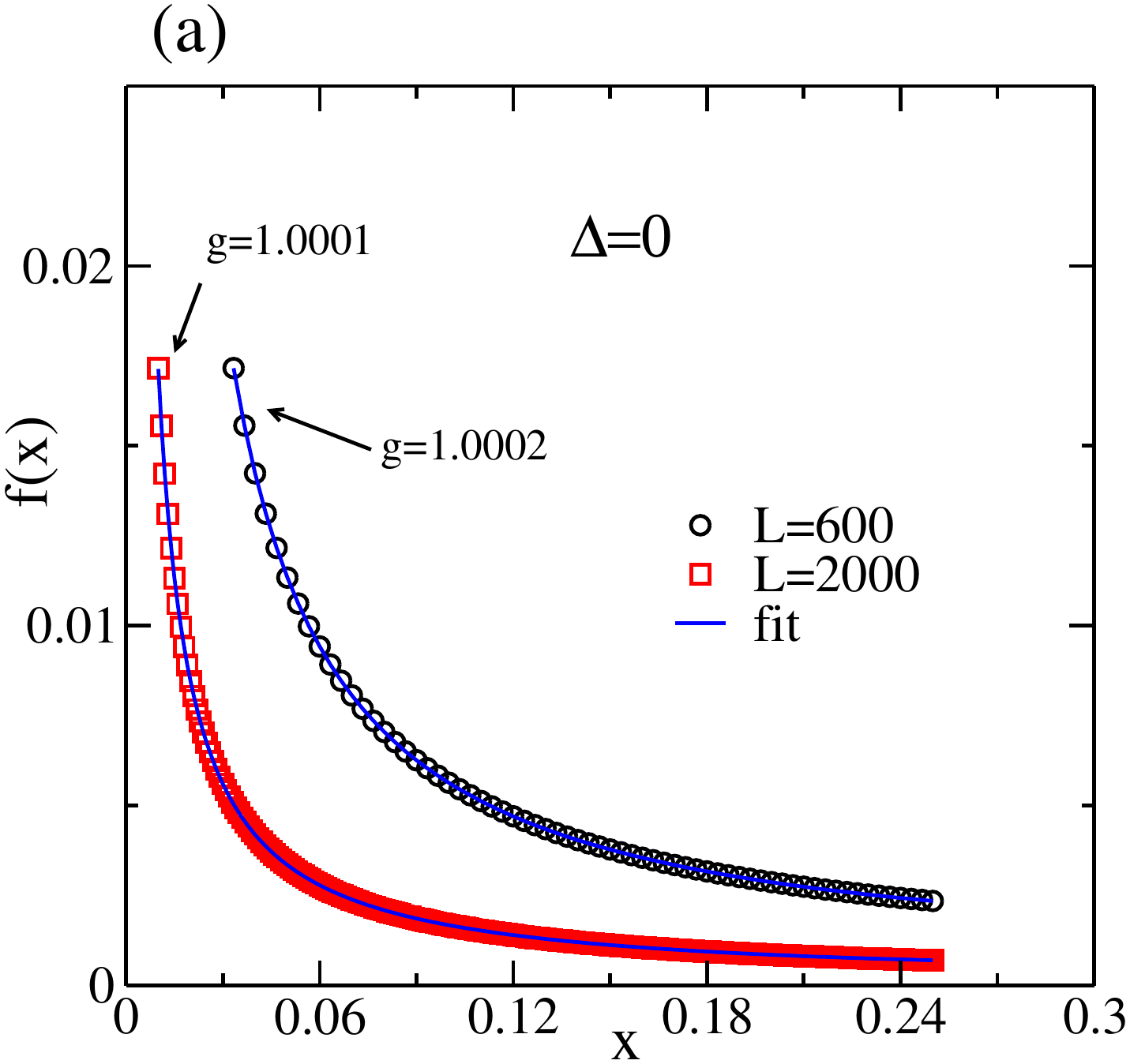} 

\includegraphics[scale=0.45]{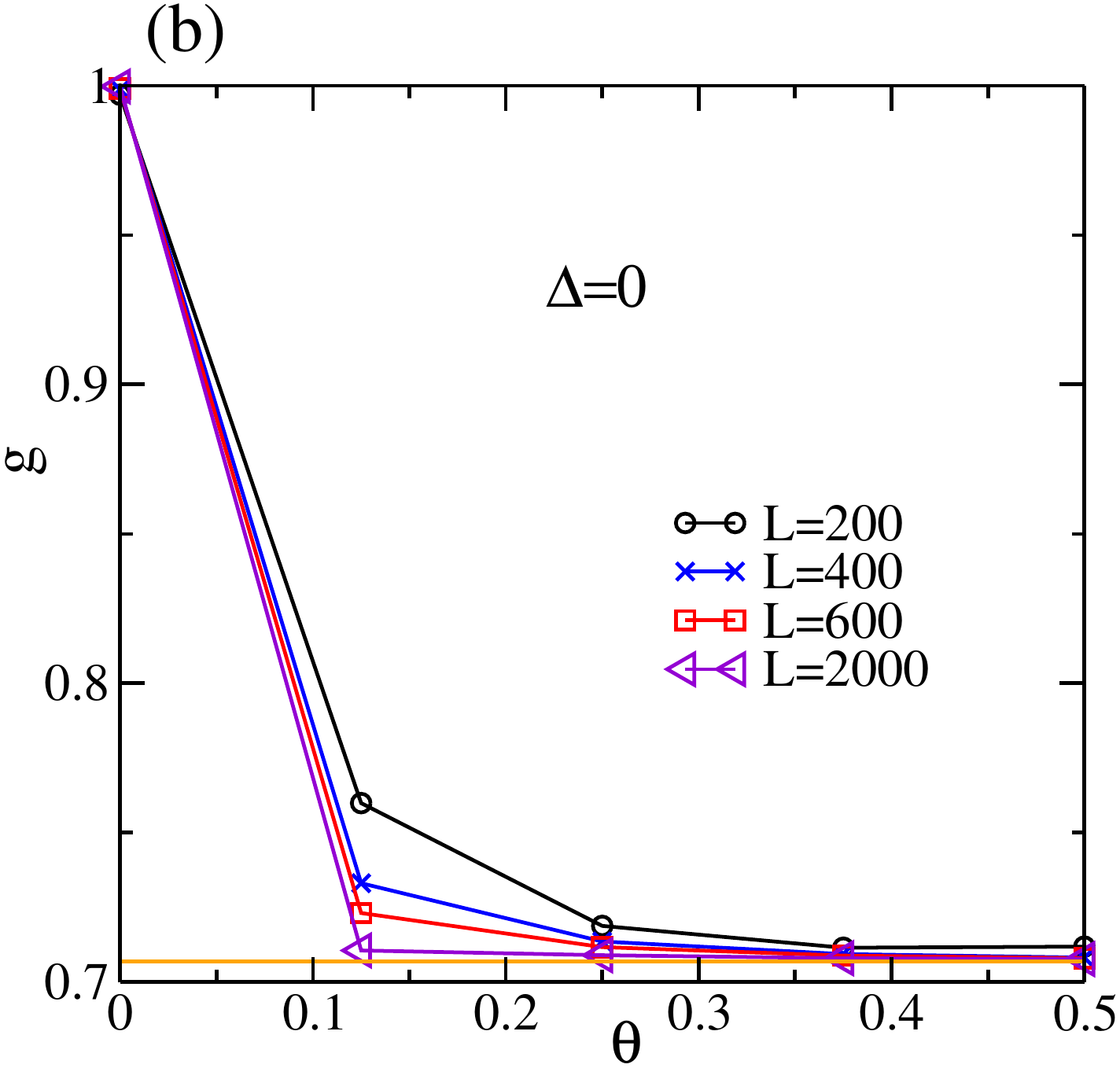}

\includegraphics[scale=0.45]{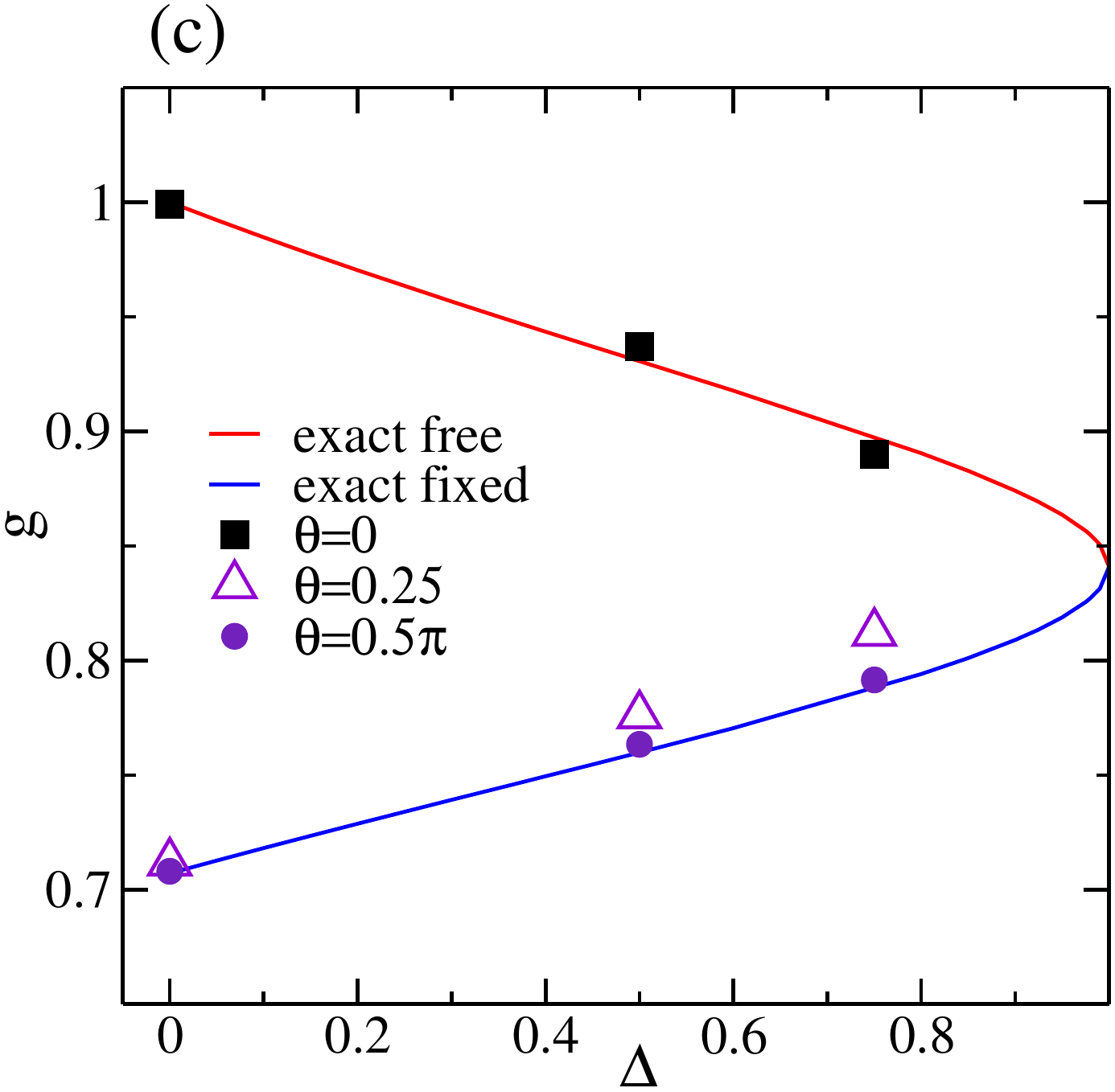}

\caption{\label{figXX} (a) The function $f(x)$ vs. $x$ for XX chain and
two system sizes (see legends). The symbols are the numerical data
and the continuous lines are the best fit to Eq. (\ref{fx}). The
arrows indicate the values of $g$ we get by the fitting procedure.
(b) Estimates of $g$ for the XX chain with $b_{1}=b_{L}=1$ and for
some values of $\theta$ and $L$. The orange solid line correspond to $g=\sqrt{2}/2\equiv g^{fixed}$. 
(c) Values of $g$ for the XXZ
chain for some boundary conditions (see text). The solid lines are
the bosonization predictions while the symbols are numerical estimates obtained considering
systems with sizes $L=600$. }
\end{figure}

We first consider the XX chain with OBC. In Fig. \ref{figXX}(a), we show the function $f(x)$ defined in Eq. (\ref{fx}) for the XX chain with OBC. By fitting the numerical data to this equation
we get $g=1.0002$ and $g=1.0001$ for $L=600$ and $L=2000$, respectively.
Similar agreement with the bosonization prediction is found also for the XXZ under
OBC, as depicted in Table \ref{tab1} for two other values of $\Delta$.
For $\Delta=0.5$ and $\Delta=\cos{(\pi/8)=0.9238...\,}$,  we used the DMRG
to obtain the entanglement entropy. For the systems under PBC (OBC and ADBMF) we kept up to $m=3000$ ($m=800$) states per block in the final sweep and done $\sim6-8$ sweeps. The discarded weight was typically around $10^{-10}-10^{-12}$ at that final
sweep.

\begin{table}
\begin{tabular}{c|ccc}
\multicolumn{3}{c}{\hspace*{1.4cm}$\Delta$ } & \tabularnewline
\hline 
\multicolumn{1}{c}{} & $0$  & $0.5$  & $0.9238$\tabularnewline
\hline 
$g^{OBC}$ & 1.0002 & 0.9309  & 0.8784 \tabularnewline
 & (1.0000)  & (0.9306)  & (0.8694)\tabularnewline
\hline 
\multicolumn{1}{c}{} &  &  & \tabularnewline
\end{tabular}

\caption{\label{tab1}The estimates of ground state degeneracy, $g^{OBC}$,
for XXZ Heisenberg chain with OBC and $L=600$ for three values of anisotropy parameter
$\Delta$. The results in parentheses are the predicted ones (see text). }
\end{table}

Finally we consider the $XXZ$ chain with ADBMF. Here too, we first focus on the XX chain. 
In Fig. \ref{figXX}(b), we preset some values of $g$ obtained by the
fitting procedure for the XX chain with some values of $\theta$. Note that For $L=600$ and $\theta=0.5\pi$, which corresponds to the magnetic field in the $x$ direction we get $g=0.708$. In this situation, it is expected that the system corresponds to a fixed conformally invariant
boundary condition with $g^{fixed}=\sqrt{2}/2$ \cite{affleckboundaryxxz}. Indeed, our result agrees very well with  the bosonization prediction for the $x$ direction. For the other directions of the boundary magnetic fields, which to our knowledge were not considered so far in the literature, our estimates also indicate  that for $0<\theta\leq\pi/2$ we have $g=g^{fixed}=\sqrt{2}/2$, as one can observe in Fig. \ref{figXX}(b).  The crossover length $\xi^x\sim h_b^{d-1}$  associated with the boundary perturbations of the boundary magnetic fields  $x$ and $z$ directions have dimensions $d_x=2\pi R^2$ and $d_z=1$, respectively \cite{affleckboundaryxxz}. Since the boundary perturbation in the $z$ direction is marginal logarithmic corrections may appear. Note that for $\theta\sim 0$ we have $\xi^x\sim (b\theta)^{2\pi R^2-1}$ and similar to the transverse field Ising case a huge crossover length is expected close to $\theta=0$ for finite boundary magnetic fields. We also estimate $g$ for other two values of the $\Delta$ [see Fig. \ref{figXX}(c)] and different directions of magnetic  fields. We summarize in Fig. \ref{figXX}(c) all the estimates of $g$  that we obtained for the XXZ chain for different boundary conditions for system sizes $L=600$ and $b=1$. As we can observe in this figure,  our results strongly support  that $g=g^{fixed}=\pi^{1/4}\sqrt{R}$ for $0<\theta\leq\pi/2$ and
$g=g^{OBC}=\frac{1}{\pi^{1/4}\sqrt{2R}}$ for $\theta=0$.

In summary as far as the boundary magnetic field vector is in the $z$ direction, i.e. $\theta=0$ one gets free boundary condition. Introducing even a small boundary  magnetic field in the $x$ and/or $y$ direction which breaks the bulk $U(1)$ symmetry induces a fixed boundary condition.


\section{Conclusions}
  \label{sec:conc}
In this paper, we investigated entanglement entropy in open quantum critical spin chains with arbitrary boundary magnetic fields. The evaluating of the boundary entropy in such systems, in general, is not a simple task by using the thermodynamic Bethe ansataz method or the CFT approach. Here, we present a simple procedure to estimate the boundary entropy by considering the finite-size corrections of the entanglement entropies and without the knowledge of the  non-universal correction $G_b$ which is induced by the boundaries \cite{Taddia2013},  see Eqs. (\ref{EEb})-(\ref{fx}).   
In particular, we calculated the boundary entropy in the critical transverse field Ising chain and the  critical XXZ chain.  We were able to obtain precise estimates of the universal boundary entropy of these two models that were in perfect agreement with previous analytical predictions. In particular, we provided estimates of the universal boundary entropy for directions of the boundary magnetic field that were not investigated in the literature so far. Our results support that if the boundary magnetic field breaks the bulk symmetry then we have a fixed boundary condition and if it does not we have a free boundary condition. 
One of our technical achievements was the exact solution of the XY chain with ADBMF which to the best of our knowledge has not been tackled so far. Our exact solution gives all the spectrum and the eigenstates of the Hamiltonian. Using this solution we were able to calculate the entanglement entropy using the modified version of the correlation method up to relatively large subsystem sizes, $L=2000$. To do similar calculations for the XXZ chain we used the DMRG.  
\newline
\newline

\textbf{Acknowledgements.}
We thank B Pozsgay for bringing in our attention to the reference [\onlinecite{Pozsgay2018}]. MAR thanks A. Jafarizadeh for discussions. MAR acknowledges partial support from CNPq and FAPERG (grant number 210.354/2018). JCX acknowledges the support from CAPES and FAPEMIG.
\newline
\newline

\appendix
\section{Calculation of  $\delta_+$}\label{SEC:delta-Plus}
In this Appendix, we show how to calculate $\delta_+=\langle\tilde{G}_{+}|\sigma^x_{L+1}|\tilde{G}_{+}\rangle$. First of all it is easy to see that 
\begin{eqnarray}\label{delta plus 1}\
\delta_+=\langle\tilde{G}_{+}|\sigma^x_{L+1}|\tilde{G}_{+}\rangle=\nonumber\hspace{4cm}\\
(-i)^{L+1}\langle\tilde{G}_{+}|a_0^y\prod_{k=1}^L a_k^xa_k^ya_{L+1}^x|\tilde{G}_{+}\rangle.\hspace{0.5cm}
\end{eqnarray} 
 Since just $a_0^x$ and $a_{L+1}^y$ depend on the $\eta_0$ and $\eta_0^{\dagger}$ one can write
\begin{eqnarray}\label{delta plus 2}\
\delta_+=(-i)^{L+1}\langle\tilde{0}|a_0^y\prod_{k=1}^L a_k^xa_k^ya_{L+1}^x|\tilde{0}\rangle.
\end{eqnarray} 
Because of the Wick's theorem one can write the above correlation as a Pfaffian as follows
\begin{eqnarray}\label{delta plus 3}\
\delta_+=(-i)^{L+1}\text{Pf}[\textbf{D}],
\end{eqnarray} 
where 
\begin{eqnarray}\label{D matrix}\
\textbf{D}= \begin{pmatrix}
     0                                  & \langle a_0^ya_1^x\rangle        & \langle a_0^ya_1^y\rangle &   ...& \langle a_0^ya_{L+1}^x\rangle \\
\langle a_1^x a_0^y\rangle              &           0                      &  \langle a_1^x a_1^y\rangle & ...&\langle a_1^x a_{L+1}^x\rangle \\
\langle a_1^ya_0^y\rangle               &  \langle a_1^ya_1^x\rangle       & 0       & ...  & \langle a_1^y a_{L+1}^x\rangle \\

.                                      &   .           &. &.                     & .\\
.                                      & .    &. &.                              &.\\
.                                      & .                                  &. & .&.   \\
\langle a_{L+1}^x a_0^y\rangle          &\langle a_{L+1}^xa_1^x\rangle      & \langle a_{L+1}^xa_1^y\rangle & ...&0
  \end{pmatrix}.
\end{eqnarray}

\section{The $\ln{2}$ term in the EE}\label{SEC:ln2}
  
For the cases that the reduced density matrix is build using an state that one site is not entanglement with the others sites, we need to be careful when we use the correlation matrix method to calculate $S(L,\ell)$. In this Appendix, we show why we should subtract the $\ln 2 $ in the entanglement entropy for a particular situation.

For simplicity, let us consider the following free fermion  Hamiltonian

\begin{equation}
H=\sum_{i,j}^L   c_{i}^{\dagger}H_{i,j}c_{j}\;.  
\end{equation}
Suppose that we project the ground state of the above Hamiltonian to obtain the state $\mid\Psi_{0}^{L}\rangle=\mid +\rangle \otimes\mid\Psi_{0}^{L-1}\rangle$, where $\mid +\rangle=\frac{1}{\sqrt{2}}\left(\mid 1\rangle+\mid 0\rangle\right)$ and $c_{1}^{\dagger}c_{1}\mid n\rangle=n\mid n\rangle$, $n=0,1$. 

Let us focus in the following density matrix 
\begin{equation}
\rho=\mid\Psi_{0}^{L}\rangle \langle\Psi_{0}^{L}\mid=\mid +\rangle\langle+\mid\otimes\mid\Psi_{0}^{L-1}\rangle \langle\Psi_{0}^{L-1}\mid\;.
\end{equation}
So, the reduced density matrix is given by
\begin{equation}
\rho_A = {\rm tr}_{B} \;  \rho \;=
(\frac{1+c_{1}^{\dagger}+c_{1}}{2})\tilde{\rho}_A\;,
\end{equation}
where we have defined the reduced density matrix associated with the sites $2,...,\ell$ as 
\begin{equation}
\tilde{\rho}_A ={\rm tr}_{B}\mid\Psi_{0}^{L-1}\rangle\langle\Psi_{0}^{L-1}\mid=
\frac{e^{-h_A}}{{\rm tr }\;e^{-h_A}}\;.
\end{equation}
We are going to assume that  $\tilde{\rho}_A$  can be written in a diagonal form in terms of new creation/annihilation operator as 
\begin{equation}
\tilde{\rho}_A=\frac{e^{{-\sum_{k=2}^\ell\epsilon_kd_{k}^{\dagger}d_{k}}}}{{\rm tr }\;e^{-h_A}}\;.
\end{equation}
Due to this fact, the eigenvalues $\epsilon_k$ are associated with the eigenvalues $\lambda_k$ of the correlation matrix $C_{i,j}={\rm tr}_{A} \left(\rho_A c_{i}^{\dagger}c_{j}\right)$, with $i,j=2,...,\ell$ by $\lambda_k=(1+e^{\epsilon_k})^{-1}\equiv \frac{1-\nu_k}{2}$ \cite{peschel2003calculation}. And the entanglement entropy for that kind of state is given by \cite{peschel2003calculation}
\begin{widetext}
\begin{eqnarray}
S(L,\ell)=-\sum_{k=2}^{\ell}[\frac{1+\nu_{k}}{2}\ln\frac{1+\nu_{k}}{2}+\frac{1-\nu_{k}}{2}\ln\frac{1-\nu_{k}}{2}]\;,\ell=2,...L\, ,
\end{eqnarray}
\end{widetext}
and for $\ell=1$ we have that S(L,1)=0.

Now, suppose that instead of considering the correlation matrix $C_{i,j}$ we define the following correlation matrix $\tilde{C}_{i,j}={\rm tr}_{A} \left(\rho_A c_{i}^{\dagger}c_{j}\right)$, with $i,j=1,2,...,\ell$. It is simple to show that  $\tilde{C}_{1,j}=1/2 \;\delta_{1,j}$. Due to this fact, the eigenvalues of the matrix  $\tilde{C}$ are the same eigenvalues as $C$ plus the eigenvalue $\lambda_1=1/2$ (which correspond to $\nu_1=0$). So, we see that if  we associate the entanglement entropy $\tilde{S}$ with the eigenvalues of the correlation matrix $\tilde{C}$, we realize  that $\tilde{S}(L,\ell)=\ln 2+S(L,\ell)$.

\bibliography{BoundaryEnt.bib}

%
%
%




\end{document}